\documentclass[conference]{IEEEtran}

\usepackage{pdfpages}
\usepackage{xcolor}
\usepackage{cite}
\usepackage{amsmath,amssymb,amsfonts}
\usepackage{algorithmic}
\usepackage{graphicx}
\usepackage{textcomp}
\usepackage{xcolor}
\usepackage[hyphens]{url}
\usepackage{fancyhdr}
\usepackage{comment}
\usepackage{tcolorbox}
\usepackage{multirow}
\usepackage{soul}
\usepackage[inkscapelatex=false]{svg}
\usepackage{xfrac}
\usepackage{cite}
\usepackage{amsmath,amssymb,amsfonts}
\usepackage{algorithmic}
\usepackage{graphicx}
\usepackage{textcomp}
\usepackage{xcolor}
\usepackage[hyphens]{url}
\usepackage{fancyhdr}
\usepackage{mathtools}
\usepackage{tikz}
\usepackage{xcolor}
\newcommand*\circled[1]{\tikz[baseline=(char.base)]{
            \node[shape=circle,fill,inner sep=2pt] (char) {\textcolor{white}{#1}};}}

\usepackage[bookmarks=true,breaklinks=true,letterpaper=true,colorlinks,citecolor=blue,linkcolor=blue,urlcolor=blue]{hyperref}

\newcommand{\ignore}[1]{}

\newcommand{\PRHTignore}[1]{}

\newcommand{\revignore}[1]{}

\newcommand{\new}[1]{{#1}}

\newcommand{\bld}[1]{{\bf{#1}}}

\definecolor{aliceblue}{rgb}{0.94, 0.97, 1.0}
\usepackage{xcolor}
\usepackage{tcolorbox}
\tcbset{width=1\columnwidth,boxrule=1pt,colback=aliceblue,arc=1pt,left=2pt,right=2pt,boxsep=0.5pt}

\title{MINT: Securely Mitigating Rowhammer\\ with a Minimalist In-DRAM Tracker \vspace{-0.15 in} }
\author{\IEEEauthorblockN{Moinuddin Qureshi}
\IEEEauthorblockA{
\textit{Georgia Tech}\\
moin@gatech.edu }
\vspace{-0.2 in}
\and
\IEEEauthorblockN{Salman Qazi}
\IEEEauthorblockA{
\textit{Google}\\
sqazi@google.com}
\vspace{-0.2 in}
\and
\IEEEauthorblockN{Aamer Jaleel}
\IEEEauthorblockA{
\textit{NVIDIA}\\
ajaleel@nvidia.com}
\vspace{-0.2 in}
}

\begin{document}
\maketitle
\pagestyle{plain}

\ignore{

This paper studies in-DRAM trackers for rowhammer mitigation

Advantage, but challenges, low cost and limited by REF

Counter based expensive cannot afford.  PRHT.

Ideally, we seek an ultra-lost cost tracker that is competitive.

In this work, we make two observations.  If we can REF one, we need only one entry.
Second, prior trackers based on past behavior or only on current access (probabilistic). We propose MINT, a single-entry tracker that makes decisions for the future at each REF.
It is both low-cost and constraints the attacker to not use more than a single copy of the address

We determine threshold -- it beats at 500-entry tracker, and has threshold competitive (within 2x) of an idealized PRHT. 

We also analyze transitive attacks and refresh postponement

We also combine with RFM to obtain thresholds of 

}

\begin{abstract}

This paper investigates secure low-cost in-DRAM trackers for mitigating Rowhammer (RH). In-DRAM solutions have the advantage that they can solve the RH problem within the DRAM chip, without relying on other parts of the system.  However, in-DRAM mitigation suffers from two key challenges: First, the mitigations are synchronized with refresh, which means we cannot mitigate at arbitrary times.  Second, the SRAM area available for aggressor tracking is severely limited, to only a few bytes. Existing low-cost in-DRAM trackers (such as TRR) have been broken by well-crafted access patterns, whereas prior counter-based schemes require impractical overheads of hundreds or thousands of entries per bank. The goal of our paper is to develop an ultra low-cost secure in-DRAM tracker.

Our solution is based on a simple observation: if only one row can be  mitigated at refresh, then we should ideally need to track only one row. We propose a {\em Minimalist In-DRAM Tracker (MINT)}, which provides secure mitigation with just a single entry. Unlike prior trackers that decide the row to be mitigated based on the past behavior (select based on activation counts) or solely based on the current activation (select with some probability), MINT decides which row in the future will get mitigated.  At each refresh, MINT probabilistically decides which activation in the upcoming interval will be selected for mitigation at the next refresh. MINT provides guaranteed protection against classic single and double-sided attacks. We also derive the minimum RH threshold (MinTRH) tolerated by MINT across all patterns. MINT has a MinTRH of 1482 which can be lowered to 356 with RFM.  The MinTRH of MINT is lower than a prior counter-based design with 677 entries per bank, and is within 2x of the MinTRH of an idealized design that stores one-counter-per-row.  We also analyze the impact of refresh postponement on the MinTRH of low-cost in-DRAM trackers, and propose an efficient solution to make such trackers compatible with refresh postponement.


\end{abstract}
\ignore{
RH, thresholds are reducing

In-DRAM trackers ...

}


\section{Introduction}

Rowhammer (RH) is a data-disturbance error that occurs when rapid activations of a DRAM row causes bit-flips in neighboring rows~\cite{kim2014flipping}. Rowhammer is a serious security threat, as it gives an attacker a powerful tool to flip bits in protected data structures, such as  page tables, which can result in privilege escalation~\cite{seaborn2015exploiting, frigo2020trrespass, gruss2018another, aweke2016anvil, cojocar2019eccploit, gruss2016rhjs, vanderveen2016drammer}  and breach of confidentiality~\cite{kwong2020rambleed}.  The RH problem has been difficult to solve, because the {\em Rowhammer Threshold (TRH)}, which is the number of activations required to induce a bit-flip, has continued to reduce with successive DRAM generations, reducing from 140K~\cite{kim2014flipping} to about 4.8K~\cite{kim2020revisitingRH} in the last decade. Thus, Rowhammer solutions must scale to low TRH.  

Typical hardware-based mitigation for RH relies on a \textit{tracking} mechanism to identify the aggressor rows (i.e., the rows that get activated repeatedly) and issue a refresh to the neighboring victim rows~\cite{hassan2021UTRR}. Hardware-based RH mitigation can be deployed at the Memory-Controller (MC) or within the DRAM chip (in-DRAM). The in-DRAM approach has the advantage that it can solve the RH problem transparently within the DRAM chip without relying on changes to other parts of the system. The in-DRAM approach also has the advantage that memory vendors can tune their mitigation solution to target the TRH observed in their chips.  This work focuses on a low-cost in-DRAM RH mitigation.  


In-DRAM RH mitigation suffers from two key constraints. {\bf First}, the RH mitigation is performed transparently within the refresh (REF) operation.  For example, for DDR5, the REF operation occurs every 3.9 microseconds, and the DRAM chip can steal some of the time reserved for REF for performing RH mitigation (refresh the victim rows of a selected aggressor row). It is not possible for the DRAM chips to sometimes take longer to do REF if there are more aggressor rows, as this would violate the deterministic timing guarantees of DDR5.  This restriction means that it is not enough to track aggressor rows, but we must {\em spread} the mitigation of aggressor rows over as many REF periods, as we cannot support {\em bursty} mitigation.  {\bf Second}, the SRAM budget available for tracking aggressor rows is quite small (often few bytes per bank) and this budget is insufficient for tracking all the aggressor rows. For example, DDR4 devices contain {\em Targeted Row Refresh (TRR)} tracker containing 1-30 entries~\cite{hassan2021UTRR}, however, such trackers have been defeated with attack patterns such as TRRespass~\cite{frigo2020trrespass} and Blacksmith~\cite{jattke2021blacksmith}. Thus, systems continue to be prone to RH attacks even in the presence of such deployed mitigations.

\begin{figure*}[!htb]
    \centering
\includegraphics[width=5.8in]{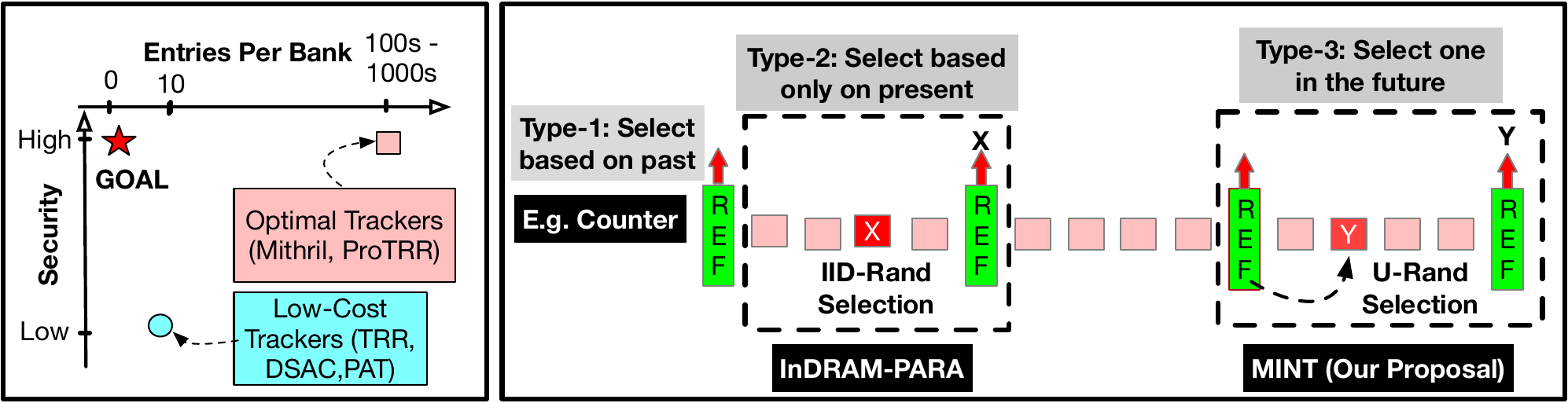}
  \vspace{-0.1 in}
    \caption{(a) Our goal is to develop secure low-cost trackers. (b) In-DRAM RH mitigation is performed at REF and the tracker decides which row to mitigate. Trackers can be categorized into three types (1) past-centric, such as counter-based tracking (2) present-centric, such as selecting the currently activated row with some probability (3) future-centric, our design, which decides at each REF which row  will be picked for mitigation in the upcoming interval. 
    }
  \vspace{-0.15 in}
    \label{fig:intro}
\end{figure*}
 
Designing secure low-cost in-DRAM trackers has proven to be a significant challenge. The recent solutions from industry have not proved to be secure either.  For example, the DSAC~\cite{DSAC} tracker from Samsung is vulnerable to Blacksmith patterns (when designed for a TRH of 2K, DSAC results in 9K unmitigated activations on an aggressor row with Blacksmith). Similarly, the PAT~\cite{HynixRH} tracker from Hynix claims 30\% lower failure rate than TRR (however, as TRR can be broken within a few minutes, PAT can also be broken within a few minutes).

Prior studies, such as ProTRR~\cite{ProTRR} and Mithril~\cite{kim2022mithril}, bound the minimum number of tracking entries required to deterministically and securely mitigate a given TRH. These studies show that an {\em optimal} number of entries in the tracker would be several thousand per bank for current TRH (e.g. 1400 entries for TRH of 2K). This SRAM overhead is prohibitively large for practical adoption in DRAM chips. Any tracker with number of entries below the optimal will always suffer from a non-zero probability of failure~\cite{ProTRR}.

\PRHTignore{
The inability to securely mitigate Rowhammer with low-cost in-DRAM trackers has caused the DRAM industry to consider radical changes~\cite{bennett2021panopticon,HynixRH}. For example, a recent design from Hynix changes the DRAM array to support {\em Per-Row Hammer Tracking (PRHT)}~\cite{HynixRH}, which incurs significant area overheads (9\%~\cite{HynixRH}) and which also requires write operation after read, resulting in performance overheads.}

Figure~\ref{fig:intro} (a) provides an overview of the landscape of in-DRAM trackers. Current in-DRAM trackers are either insecure or incur a prohibitively high-cost. The goal of our paper is to develop a low-cost secure in-DRAM tracker. 
Our solution is based on a simple insight:  If in-DRAM mitigation is limited to mitigating at-most one aggressor row per REF, ideally we should need a tracker with only a single-entry (that identifies the aggressor row to be mitigated). We develop a classification of in-DRAM trackers that helps in guiding our solution.  Figure~\ref{fig:intro} (b) shows an overview of in-DRAM rowhammer mitigation. A REF occurs at each tREFI interval and there are activations between REF.  At each REF, one aggressor row is selected for mitigation. Depending on how this selection is made, we can classify the in-DRAM trackers into three types.  First, {\bf past-centric}, which makes the selection decision based on past behavior.  For example, counter-based schemes (ProTRR and Mithril) select a row with the maximum counter value. Such trackers need a large amount of storage.  Second, {\bf present-centric}, which makes the {\em selection} decision probabilistically based only on the currently activated row.  For example, a PARA~\cite{kim2014flipping} like in-DRAM scheme would select each activated row with an {\em Independent and Identically Distributed (IID) } probability of $p$. If selected, the row address is stored in a single-entry tracker for getting mitigated at the next REF. The problem with such an {\em {\textcolor{black}{InDRAM-PARA}}} design is that a selected row can be {\em over-written} before reaching REF by another selected row and thus miss the chance of mitigation.  Furthermore, even if the same row gets activated throughout the tREFI interval, there is still a non-negligible likelihood that the row will still not be selected for mitigation.  Ideally, we want a single-entry tracker that avoids both problems of \textcolor{black}{InDRAM-PARA} (no over-writing a selected entry and guaranteed selection of one row in tREFI). The {\em Minimum TRH (MinTRH)} tolerated by the \textcolor{black}{\textcolor{black}{InDRAM-PARA}} is 7.6K. 

We propose a {\em Minimalist In-DRAM Tracker (MINT)} that provides secure RH mitigation with a single-entry. MINT offers a new category, the third-type, {\bf future-centric}. Let there be a maximum of $M$ activations in tREFI. At each REF, MINT decides which of the M activations in the {\em upcoming} interval should be selected for mitigation at the {\em next} REF. MINT performs this selection using a {\em Uniform Random (U-RAND)} choice of all 1 to M positions. This position is stored in a {\em Selected Activation Number (SAN)}. Each activation in tREFI is given a sequence number, and when this number reaches SAN, the row is designated for mitigation at  next REF. 

By design, MINT avoids over-writing the selection. Secondly, MINT guarantees selection of one row (from M activations), thus it provides guaranteed protection against classic single-sided and double-sided attacks, if such attacks are done continuously. We analyze the worst-case pattern for MINT for determining the MinTRH.  We also analyze {\em Transitive Attacks}~\cite{HalfDouble} that leverage victim refresh to cause RH failures in a distant row.  Our analysis shows that MINT has a  MinTRH of 2800 (MinTRH-D of 1400 for double-sided). The MinTRH of MINT is similar to Mithril~\cite{kim2022mithril} with 677 entries per bank. 

DDR5 allows the postponement of up-to four REF operations. Delayed refreshes are especially problematic for low-cost trackers as they track limited entries, which may get dislodged (without mitigation). We analyze the impact of delayed refreshes on low-cost trackers and propose a generalized solution, the {\em Delayed Mitigation Queue (DMQ)}, that enable low-cost trackers to operate with REF postponement. With DMQ, the MinTRH-D of MINT is 1482. 


\begin{tcolorbox}
The threshold tolerated by MINT (with support for refresh postponement) is lower than a Mithril tracker with 677-entries per bank, and is within {\bf{2x}} of an idealized tracker that has one counter per row. Thus, our proposal bounds the gap between the lowest-cost tracker (single-entry) and an idealized tracker (per-row counters) to a narrow range (2x). 
\end{tcolorbox}

Overall, this paper makes the following contributions:
\vspace{0.05 in}

\begin{enumerate}
    \item We propose {\em Minimalist In-DRAM Tracker (MINT)} that provides secure RH mitigation with a single entry.
 \vspace{0.05 in}

    \item We show that MINT tolerates a MinTRH of 2800 (MinTRH-D  1400), similar to a 677-entry tracker.
\vspace{0.05 in}

    \item To the best of our knowledge, this is the first paper to study the impact of refresh postponement on low-cost trackers.  We propose {\em Delayed Mitigation Queue (DMQ)} to enable low-cost tracking with refresh  postponement. MINT+DMQ has a MinTRH-D of 1482. 
\vspace{0.05 in}
    
    \item We combine MINT with the RFM feature of DDR5 to to obtain a MinTRH-D as low as 356. 
  \vspace{0.05 in}
  
\end{enumerate}

The storage overhead of MINT is four bytes (per bank) and the performance and power overheads are negligibly small.  
\clearpage

\ignore{
 DRAM background

 Threat Model

 Rowhammer and TRH

 MC-Based Mitigation

 In-DRAM Mitigation

 Low Cost Trackers

 Optimal Trackers

 Goal
}


\section{Background and Motivation}

\subsection{DRAM Architecture and Parameters.}

To access data from DRAM, the memory controller must first issue an activation (ACT) for the DRAM row. To ensure data retention, the memory controller sends a refresh command every tREFI that refreshes a subset of rows. Table~\ref{table:Params} shows the DDR5 parameters, derived from  DDR5 datasheet (DDR5-5200B bin with 32Gb chips). The two critical parameters for our study are: (1) The maximum number of ACT (MaxACT) possible within tREFI is 73 and (2) We assume the device performs one Rowhammer mitigation at each refresh event.

\begin{table}[!htb]
  \centering
  \vspace{-0.1in}
  \caption{DRAM Parameters (from DDR5 datasheet)}
  \vspace{-0.1in}
  \begin{footnotesize}
  \label{table:Params}
  \begin{tabular}{lcc}
    \hline
    \textbf{Parameter} & \textbf{Explanation} & \textbf{Value} \\ \hline
    \rule{0pt}{0.8\normalbaselineskip}tREFW     & Refresh Window & 32 ms \\ 
    tREFI     & Time interval between REF Commands & 3900 ns  \\ 
    tRFC      & Execution Time for REF Command & 410 ns  \\ 
    tRC       & Time between successive ACTs to a bank & 48 ns \\ \hline 
    
     MaxACT & M = ( tREFI - tRFC )~/~tRC & 73 \\ \hline
     
\hline 
  \end{tabular}
  \vspace{-0.1in}
  \end{footnotesize}
\end{table}

\subsection{Threat Model}
Our threat model assumes an attacker can issue memory requests for arbitrary addresses. 
We assume the attacker knows the defense algorithm, but does not have physical access to the system (e.g., outcome of random-number generator).
Our defense aims to prevent Rowhammer against all access patterns, including Blacksmith~\cite{jattke2021blacksmith}, and Half-Double~\cite{HalfDouble}.  To keep our analysis simple, we do not consider the Row-Press~\cite{rowpress} attack. Our recent work~\cite{IMPRESS}  shows that Row-Press can be easily mitigated by converting the row-open time into an equivalent number of activations (Appendix C shows how our solution can be modified to tolerate Row-Press).

\subsection{DRAM Rowhammer Attacks}
Rowhammer~\cite{kim2014flipping} occurs when a row (aggressor) is activated frequently, causing bit-flips in nearby rows (victim). The minimum number of activations to an aggressor row to cause a bit-flip in a victim row is called the {\em Rowhammer Threshold (TRH)}. TRH can be reported either for a single-sided pattern {\em (TRH-S)} or a double-sided pattern {\em (TRH-D)}. As shown in Table~\ref{table:TRH},  TRH has dropped significantly, from 139K (TRH-S) in 2014~\cite{kim2014flipping} to 4.8K (TRH-D) in 2020~\cite{kim2020revisitingRH}.


\begin{table}[htb]
  \centering
  \begin{footnotesize}
    \vspace{-0.1in}
  \caption{Rowhammer Threshold Over Time}
  \vspace{-0.1 in}
  \label{table:TRH}
  \begin{tabular}{ccc}
    \hline
   \textbf{DRAM Generation} & \textbf{TRH-S (Single-Sided)} & \textbf{TRH-D (Double-Sided)}\\ \hline 
    
    \rule{0pt}{0.8\normalbaselineskip}{DDR3-old}     & 139K~\cite{kim2014flipping}  & -- \\ 
    \rule{0pt}{0.8\normalbaselineskip}{DDR3-new}     &  --  & 22.4K~\cite{kim2020revisitingRH} \\ 
   {DDR4}  &  -- & 10K~\cite{kim2020revisitingRH} - 17.5K~\cite{kim2020revisitingRH} \\ 
   {LPDDR4}  & --  & 4.8K ~\cite{kim2020revisitingRH} - 9K~\cite{HalfDouble} \\ \hline   
  \end{tabular}
  \vspace{-0.05in}
  \end{footnotesize}
\end{table}

Rowhammer is  a serious security threat, as an attacker can use it to flip bits in the page-table to perform privilege escalation~\cite{seaborn2015exploiting,gruss2018another,frigo2020trrespass,zhang2020pthammer} or break confidentiality~\cite{kwong2020rambleed}. 

Solutions for mitigating Rowhammer typically rely on a mechanism to identify the aggressor rows and then performing a mitigation by refreshing the victim rows. The identification of aggressor rows can be done either  at the Memory Controller (MC) or within the DRAM chip (in-DRAM).

\subsection{Memory-Controller Based  Mitigation}

Memory-Contoller (MC) based solutions identify aggressor rows either using counters~\cite{lee2019twice}\cite{park2020graphene}\cite{qureshi2022hydra} or probabilistically~\cite{kim2014architectural}\cite{kim2014flipping}. 
 These solutions suffer from three major shortcomings. First, DRAM chips internally use proprietary mapping, so such a solution must rely on the {\em Directed RFM (DRFM)}  command of DDR5 to do mitigation.  DRFM incurs a latency of tRFC (410ns), resulting in significant slowdown, and there is a rate limit of one DRFM per two tREFI, placing a high limit on the TRH that can be tolerated. Second, the solution must be conservatively designed for the lowest TRH, across vendors and over the years of the system lifetime.  Third, the cost and complexity of tracking can deter some processor vendors from adoption, leading to fragmented protection.
 

\subsection{In-DRAM Mitigation}

The advantage of in-DRAM mitigation is that it can solve Rowhammer within the DRAM chips, without relying on other parts of the system.  Furthermore, DRAM manufacturers can tune their solution to the TRH of their chips.

In-DRAM mitigation typically performs the mitigation transparently during the time provisioned for the refresh operations (commodity DRAM is a deterministic device, so it is not possible to do mitigative activations while servicing the normal demand accesses). A recent study~\cite{ProTRR} observes that DDR5 chips support mitigating one aggressor row at each tREFI, or one per two tREFI. In our paper, we assume a default rate of mitigating one aggressor row per tREFI.

\subsection{Low-Cost In-DRAM Trackers: Not Secure}

For guaranteed protection, the in-DRAM tracker must be able to identify \textit{all} aggressor rows and mitigate them before they receive TRH activations. Unfortunately, the SRAM budget available for tracking within the DRAM chip is limited to only a few bytes. Therefore, practical in-DRAM trackers (such as {\bf TRR} from DDR4, {\bf DSAC}~\cite{DSAC} from Samsung, and {\bf PAT}~\cite{HynixRH} from SK Hynix) are limited to only tracking a few entries (1-30), and can be broken within a few minutes using patterns that target a large number of aggressor rows~\cite{frigo2020trrespass} or use decoy rows~\cite{jattke2021blacksmith}. Thus, the system remains vulnerable to RH attacks, even in the presence of such low-cost trackers.

\ignore{
Examples of low-cost trackers include:

\vspace{0.05 in}

\noindent{\bf TRR (DDR4):} DDR4 modules perform {\em Targeted Row Refresh (TRR)}~\cite{frigo2020trrespass} using a tracker of up to 30 entries~\cite{hassan2021UTRR, jattke2021blacksmith}. RR has been broken using the \textit{TRRespass}~\cite{frigo2020trrespass} and  \textit{Blacksmith}~\cite{jattke2021blacksmith}.

\vspace{0.05 in}

\noindent{\bf DSAC (Samsung)~\cite{DSAC}:} This is a recent proposal from Samsung that uses 20 entries each with a counter. Accessed  row replaces the min-counter entry with a probability inversely proportional to min-counter+1. DSAC is vulnerable to Blacksmith.

\vspace{0.05 in}

\noindent{\bf PAT (SK Hynix)~\cite{HynixRH}:} This recent proposal from SK Hynix uses an in-DRAM tracker with 8 entries. The paper reports that PAT has 30\% lower failure rate than conventional designs, which means that PAT can be broken with a few minutes. 

\vspace{0.05 in}
}


\subsection{Optimal In-DRAM Trackers: Not Practical}

The minimum number of entries needed for an in-DRAM tracker to deterministically and securely tolerate a threshold of TRH is determined by the rate of mitigation (e.g. one per tREFI). If several rows get identified as aggressor rows at a similar time, then in-DRAM solution would need to spread their mitigation over several tREFI intervals, leading to more activations on the unmitigated aggressor rows. Two concurrent works~\cite{kim2022mithril}~\cite{ProTRR}, establish the bounds on TRH tolerated by in-DRAM mitigation arising from such a restriction.  They also bound the optimal number of entries needed to tolerate a given TRH. We call the designs that have the minimum number of tracking entries to tolerate a given threshold as {\em optimal} trackers. Examples of optimal trackers include:

\vspace{0.05 in}

\noindent{\bf Mithril~\cite{kim2022mithril}:}  Mithril uses a {\em Counter-based Summary} algorithm to track the activation counts of heavily activated rows.  At mitigation, the row with the highest counter value is mitigated and the counter value is reduced by the min count.

\vspace{0.05 in}

\noindent{\bf ProTRR~\cite{ProTRR}:} ProTRR performs {\em victim tracking} using a {\em Misra-Gries} tracker to identify the top victim rows.  At mitigation, the victim row(s) with the highest counter value get refreshed and removed from the tracker.   

Given the TRH and the rate of mitigation, we can determine the number of entries in the optimal trackers. For example, for a mitigation rate of 1 per tREFI, for a TRH-D of 1K, Mithril would require approximately 1400 entries.  Note that each bank requires an independent tracker, so the total number of tracker entries to protect the entire DRAM rank (32 banks) would be approximately 45K.  Unfortunately, the limited SRAM budget within the DRAM renders such trackers impractical.

\PRHTignore{
\subsection{Per-Row Counter-Table (PRCT): An Idealized Design}
}
\subsection{Per-Row Counter-Table (PRCT)}

\PRHTignore{
The inability to design a low-cost secure tracker based on the limited SRAM available within the DRAM chip is causing some companies to consider tracking using the DRAM arrays~\cite{bennett2021panopticon,HynixRH}. For example, a recent design from Hynix changes the DRAM array to include a counter with each DRAM row to track the number of {\em hammers} (activations).  Such a {\bf{Per-Row Hammer Tracking (PRHT)}}~\cite{HynixRH} offers an idealized design as each row has a dedicated counter.  However, this design suffers from significant area overheads (reported as 9\%~\cite{HynixRH}). It also requires changing the activation command to support a write after each activation, resulting in performance impact. Finally, the threshold tolerated by this design is still limited by the rate of mitigation. For example, at 1 mitigation per tREFI, PRHT can let two aggressor rows reach 623 activations each (victim subjected to 1226 activations). Thus, the MinTRH tolerated by PRTH is 1226 (623 double-sided). 
}


The TRH tolerated by a in-DRAM mitigation depends on both, the number of entries in the tracker and the rate of mitigation (e.g. one per tREFI).  To understand the limit of in-DRAM mitigation, under a given rate-of-mitigation, we also study an idealized design, {\em{Per-Row Counter-Table (PRCT)}}, which stores one counter per row in an SRAM table.  Given the large overheads, such a design is not practical, however, it can still help us understand the MinTRH gap between a practical design and an idealized design.  The MinTRH tolerated by PRCT is purely determined by the rate of mitigation. For example, at 1 mitigation per tREFI, PRCT can let two aggressor rows reach 623 activations each (victim subjected to 1226 activations). Thus, the MinTRH of PRCT is 1226 (623 double-sided).

\subsection{Understanding the Impact of Refresh Postponement}

DDR5 allows the postponement of up-to four REF operations. Delayed refreshes can cause a row selected for mitigation to be further subjected to an additional 292 (73x4) activations.  For counter-based mitigations~\cite{kim2022mithril,ProTRR,HynixRH} refresh postponement increases the tolerated TRH by 292. For example, the MinTRH tolerated by PRCT with refresh postponement becomes 1518 (double-sided 759). Delayed refreshes are especially challenging for low-cost trackers with only a few entries, as those entries may get dislodged before getting mitigated during the period of refresh postponement. For example, a PARFM tracker~\cite{kim2022mithril} that tolerates a threshold of a few thousand (without refresh postponement) can be made to deterministically cause {\bf 487K (!)} activations on an attack row without any mitigation under refresh postponement.

\subsection{Goal of our Paper}

The goal of our paper is to develop an ultra low-cost and secure in-DRAM tracker with a tolerable threshold that is close to the idealized design (PRCT). Furthermore, we want our proposed design to be fully compatible with  refresh postponement, as handling such a feature is a non-negotiable requirement for practical adoption. We first discuss the pitfalls of simply extending PARA to an in-DRAM setting.

\newpage

\color{black}

\section{Fundamental Pitfalls of In-DRAM-PARA}

PARA~\cite{kim2014flipping} is a memory controller scheme that mitigates each activated row with probability {\em p}. In this section, we show that applying PARA to in-DRAM setting (InDRAM-PARA) suffers from two fundamental shortcomings: (a) non-uniform mitigation probability, depending on where the row activation lies in the tREFI interval (b) frequent non-selection of any activated row in the tREFI interval even if all the activation slots are used. These shortcomings cause the threshold tolerated by InDRAM-PARA to be {\bf{2.7x higher}} than an idealized policy that mitigates all activations with equal probability.

\begin{figure}[!htb]
    \centering
\includegraphics[width= 2.5 in]{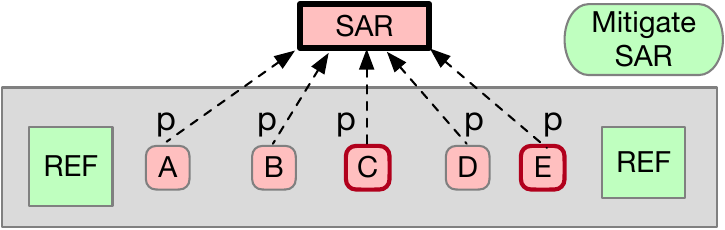}
    \caption{Design of InDRAM-PARA. Each activation is sampled with a probability p and stored in SAR. At REF the row in SAR (if valid) is mitigated.}
    \vspace{-0.15 in}
    \label{fig:iidt}
\end{figure}

\subsection{InDRAM-PARA: Design and Analysis}

Figure~\ref{fig:iidt} shows the overview of the InDRAM-PARA. Each activation is sampled with {\em Sampling Probability} {\em(p)}.  If sampled, the row-address is stored in {\em SAR (Sampled Address Register)}.  At REF, if SAR is valid, the row is mitigated.  For a row to get mitigated it must be both {\em sampled} and it must {\em survive} until REF in SAR. For example, if Row-C is sampled, then later sampling of Row-E evicts Row-C in SAR. For our design, all activations are sampled with a uniform probability (p=1/73), so the mitigation probability is proportional to the survival probability, as captured in  Equation~\ref{eq:mitig}. 

\begin{equation}
\label{eq:mitig}
P_{mitigation} = P_{sample} * P_{survive} 
\end{equation}

\vspace{0.05 in}
\noindent{\bf Model for Survival Probability:} Let there be $M$ activations between two refreshes (tREFI).  The window starts with an empty SAR. Let  Row-A be accessed at the Kth activation and get sampled into SAR. SAR will retain this entry if there is no other insertion in the remaining (M-K) activations.  If $p$ (we use $p=1/73$) is the sample probability, then the {\em Survival Probability ($S_K$)} for position $K$ is given by Equation~\ref{eq:seb}. 

\begin{equation}
\label{eq:seb}
S_K = (1-p)^{(M-K)}
\end{equation}

\begin{figure}[!htb]
    \centering
    \vspace{-0.1 in}
\includegraphics[width= 3.4 in]{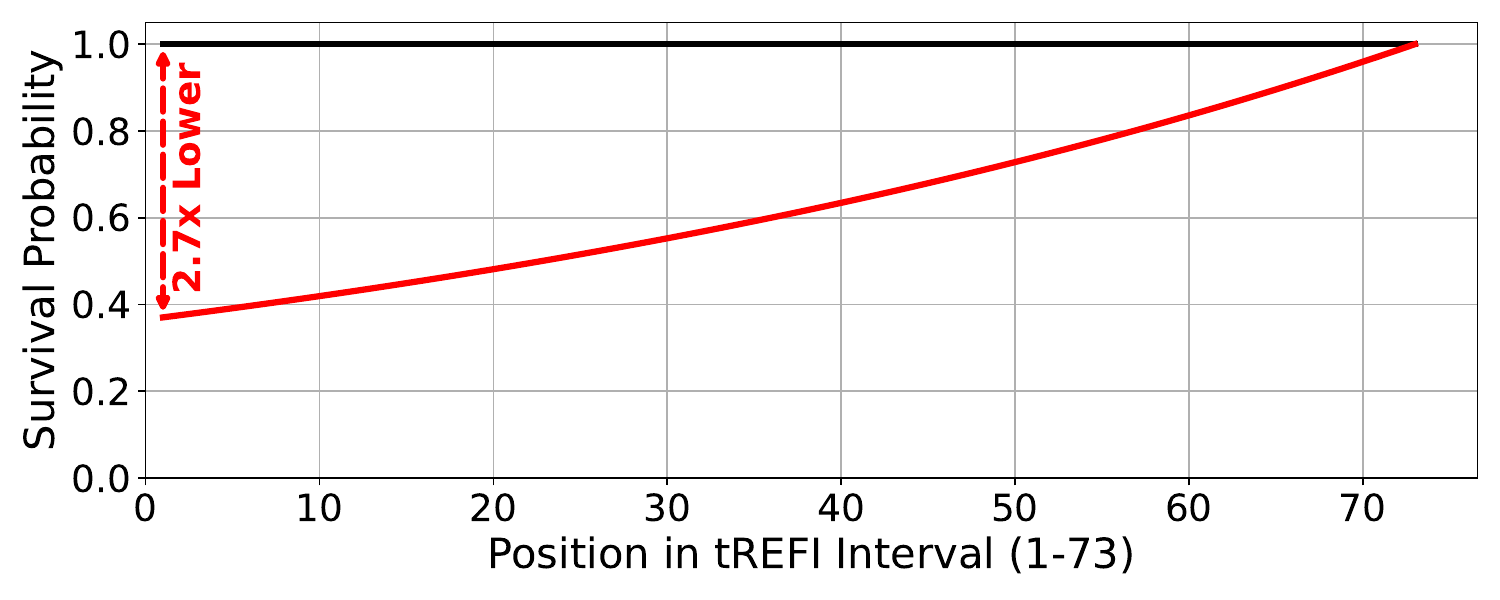}
\vspace{-0.15 in}
    \caption{Highly Non-Uniform Survival Probability for InDRAM-PARA}
    \vspace{-0.1 in}
    \label{fig:overwrite}
\end{figure}

Fig~\ref{fig:overwrite} shows the survival-probability ($S_K$) as the position (K) is varied from 1 (earliest in tREFI) to 73 (last in tREFI).

The first position has the lowest survival-probability (0.37) whereas the last position has the highest survival-probability (1). Thus, with this design, the most vulnerable position has {\bf{2.7x lower mitigation probability}} compared to an idealized scheme that mitigates all positions with probability p.

\subsection{InDRAM-PARA: No-Overwrite Version}

\begin{figure}[!htb]
    \centering
    \vspace{-0.05 in}
\includegraphics[width= 2.5 in]{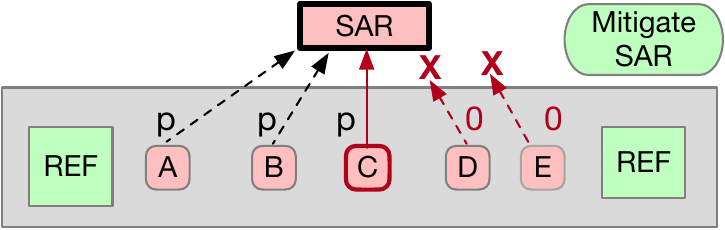}
 \vspace{-0.05 in}
    \caption{Design of InDRAM-PARA that avoids overwriting SAR. While this design has 100\% survival probability, it has non-uniform sampling probability.}
    \vspace{-0.1 in}
    \label{fig:designnooverwrite}
\end{figure}

It is intuitive to  think that the non-uniform mitigation of InDRAM-PARA can easily be addressed by simply avoiding the overwrite of SAR, if SAR had a valid entry. Figure~\ref{fig:designnooverwrite} shows an overview of such an InDRAM-PARA (No-Overwrite) design. While such a design guarantees 100\% survival probability, it suffers from another equally critical problem of non-uniform sampling. If a row gets sampled (e.g. Row-C), then the probability of sampling for all later rows (e.g. Row-D and Row-E) becomes zero. As survival probability is 1, the mitigation probability of a activation is equal to the sampling probability of that position within the tREFI interval. 

\vspace{0.05 in}
\noindent{\bf Model for Non-Uniform Sampling Probability:} Let $M$ activations between two refreshes (tREFI).  The window starts with an empty SAR. Let $p$ be the designated sampling probability at the start of the tREFI interval.  Then the first row will be selected with probability p. For all subsequent rows, the sampling probability is either p or 0, depending whether something was selected before.  The sampling probability at position K $(P_K)$ is given by Equation~\ref{eq:nooverwrite}. 

\begin{equation}
\label{eq:nooverwrite}
P_K = p \cdot (1-p)^{K}
\end{equation}

Fig~\ref{fig:overwrite} shows the sampling probability ($P_K$) as the position (K) is varied from 1 (earliest in tREFI) to 73 (last in tREFI), normalized to the first position in the window (which equals p=1/73). We note that the sampling probability is highly non-uniform, reducing to about 0.37x for the last position in the window (so absolute sampling probability has reduced from 1/73 to 1/73 * 0.37 = 1/200). Thus, even with this design, the most vulnerable position has {\bf{2.7x lower mitigation probability}} compared to an idealized scheme that performs mitigation of all positions with probability p.

\begin{figure}[!htb]
    \centering
\includegraphics[width= 3.4 in]{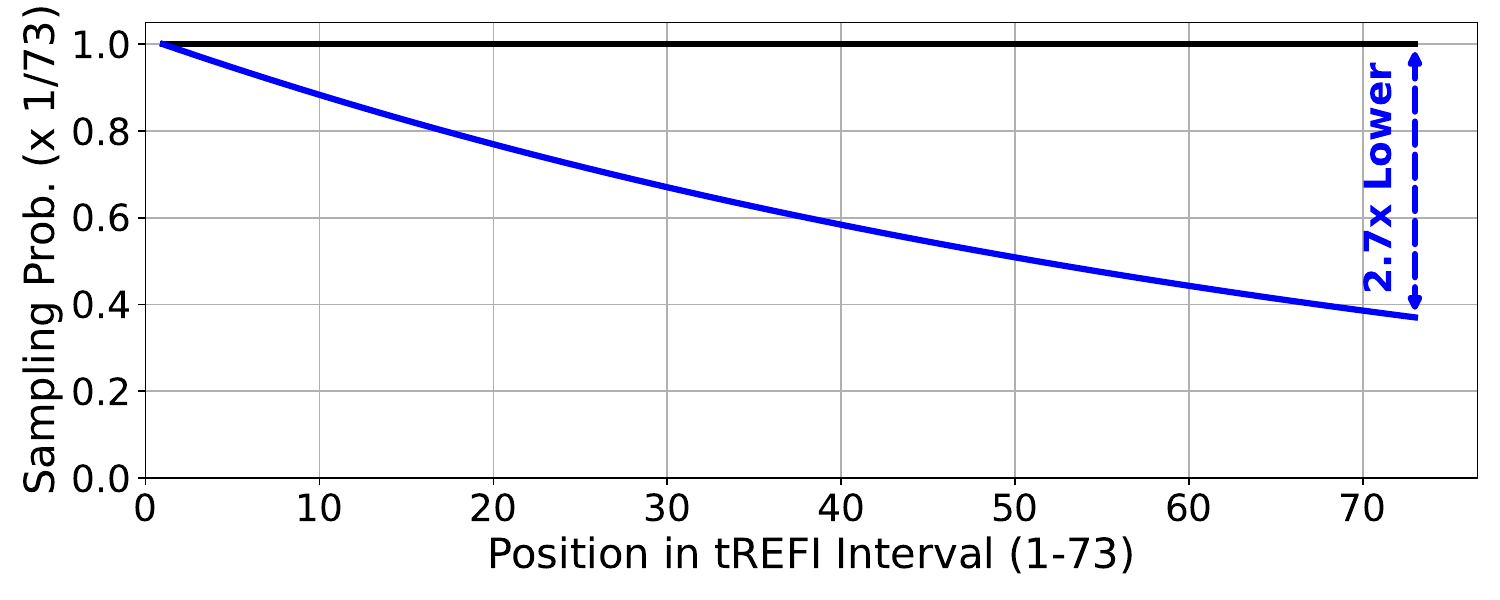}
  \vspace{-0.15 in}
    \caption{Sampling Probability for InDRAM-PARA (No-Overwrite)}
    \vspace{-0.15 in}
    \label{fig:nooverwrite}
\end{figure}

\subsection{Impact of Non-Uniform Mitigation on Security}

Figure~\ref{fig:mitprob} shows the mitigation probability of InDRAM-PARA and InDRAM-PARA(No-Overwrite) normalized to an ideal policy that mitigates each position with probability p=1/73, as the position in the tREFI window is varied.  Both versions of InDRAM-PARA Designs have non-uniform mitigation, just that the most vulnerable position is different for them (either first or last). For both designs, the most vulnerable position has 2.7x lower mitigation rate than the ideal policy.

\begin{figure}[!htb]
    \centering
     \vspace{-0.1 in}
\includegraphics[width= 3.4 in]{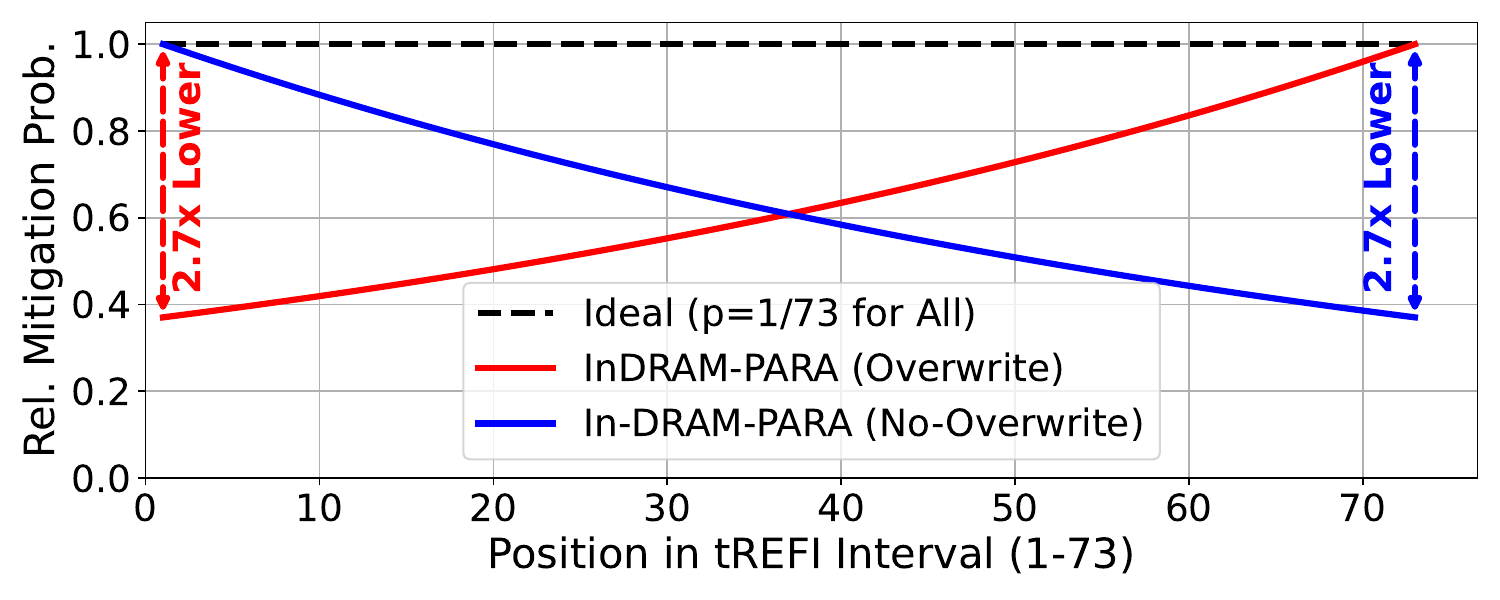}
 \vspace{-0.1 in}
    \caption{Mitigation Probability of InDRAM-PARA, InDRAM-PARA (No-Overwrite) normalized to ideal policy with uniform mitigation (p=1/73).  }
    \vspace{-0.15 in}
    \label{fig:mitprob}
\end{figure}

\vspace{0.05 in}

\noindent{\bf{Impact:}} Given any In-DRAM scheme, the attacker will focus on the most-vulnerable position, so the overall security of the design is determined by the most-vulnerable position. We note that exploiting the non-uniform mitigation probability of In-DRAM Trackers is a well-known technique in Rowhammer attacks. For example, both SMASH~\cite{de2021smash} and BlackSmith~\cite{jattke2021blacksmith} use {\em Refresh Interval Synchronization} to converge on the most vulnerable position within the tREFI window. Thus, non-uniform mitigation has significant security implications.

Given 2.7x reduced mitigation probability, the threshold tolerated by InDRAM-PARA is about 2.7x higher than an ideal policy that mitigates all positions uniformly. For example, the MinTRH of Ideal is 2.8K and InDRAM-PARA is 7.6K.

\subsection{The Problem of Non-Selection with InDRAM-PARA}

Another shortcoming of InDRAM-PARA is that even if all the activation slots of a tREFI window are used, it can still have a significant probability that nothing will get selected, thus missing out on the possibility of performing mitigation at REF.  Equation~\ref{eq:nonselection} shows the probability that nothing will be selected in a tREFI window if M activations occur.  
\begin{equation}
\vspace{-0.01 in}
\label{eq:nonselection}
P_{NoSelect} =  (1-p)^{M} = (1-1/73)^{73} =0.37
\vspace{-0.01 in}
\end{equation}

In our case, p=1/73, and M can be up-to 73.  Thus, even if all activations slots are used, InDRAM-PARA will {\bf skip mitigation 37\% of the time}. 
The non-selection of InDRAM-PARA allows stressful attack patterns, such as the classic Single-Side and Double-Sided Rowhammer attacks that continuously activate the same one or two rows over the entire tREFI.

\begin{tcolorbox}
{\bf{Key Takeaways:}} InDRAM-PARA suffers from non-uniform mitigation over the tREFI interval, and such non-uniformity has a significant impact on security and threshold. It also suffers from non-selection.  An ideal In-DRAM solution must provide uniform mitigation probability for all activations, and avoid non-selection. We propose such an ideal solution. 
\end{tcolorbox}

\ignore{

Figure~\ref{fig:mitprob} shows the probability of non-selection for both InDRAM-PARA (p=1/73, IID) and the ideal policy that mitigates all positions with equal probability (p=1/73, non-IID). If all slots are used, the ideal policy is guaranteed to pick one of the slot.

\begin{figure}[!htb]
    \centering
     \vspace{-0.1 in}
\includegraphics[width= 3.4 in]{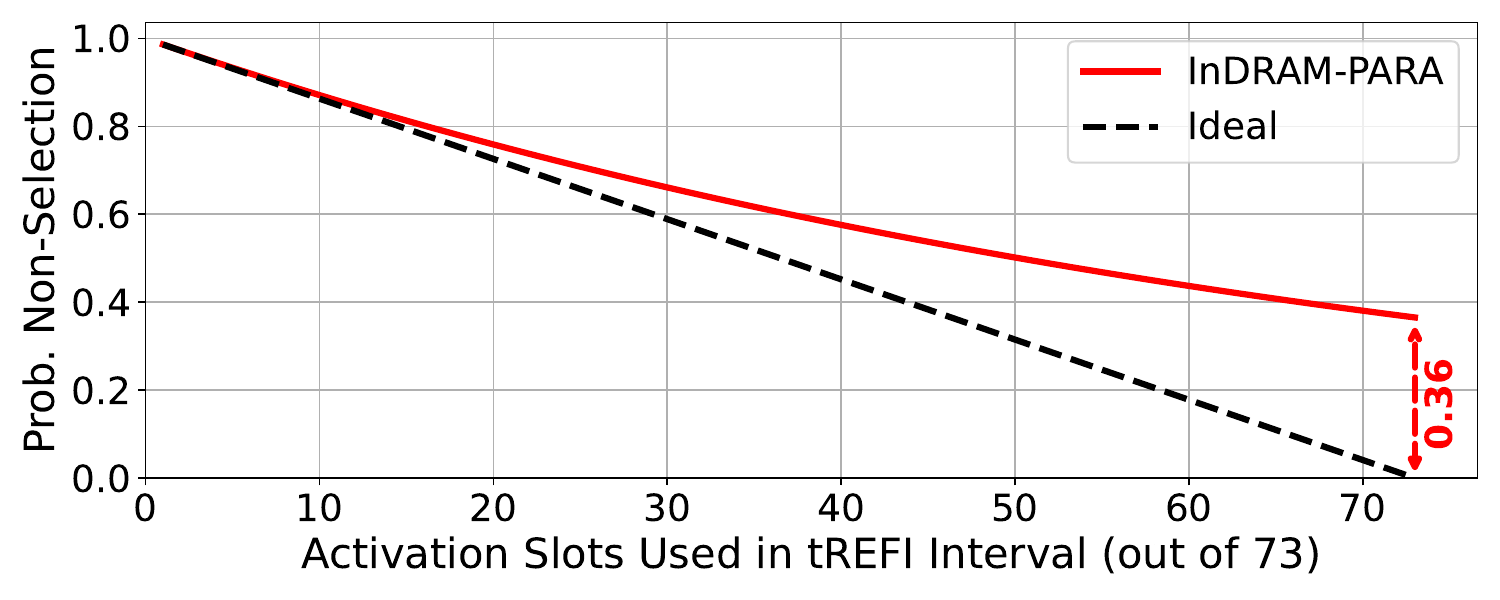}
\vspace{-0.15 in}
    \caption{Probability of non-selection as the number of activation slots in tREFI get used for InDRAM-PARA and an idealized mitigation scheme.}
    \vspace{-0.2 in}
    \label{fig:mitprob}
\end{figure}

}

\color{black}

\newpage
\section{Methodology For Analyzing Security}\label{sec:analytical_methods}

We consider an event of one or more bitflips from Rowhammer as a failure. Thus, if any row receives TRH activations, without an intervening mitigation, we declare it as a failure.  





\subsection{Model for Failure Probability in tREFW Window}

We divide the time into windows of tREFW, as all rows get refreshed every tRFW.  We want to determine the probability of failure at the $k$th activation, given the row is mitigated with probability $p$ at each activation. To the best of our knowledge, the most complete analytical model for estimating the probability of failure is by Sariou and Wolman~\cite{sariou} (it corrects an off-by-one error of Mithril~\cite{kim2022mithril} and also incorporates auto-refreshes). The probability of failure ($P_k$) at any given activation ($k$) for a TRH of $T$ is given by Equations 1-3.
\begin{equation}
\label{eq1}
    P_k = 0 \hspace{1.85 in} \textrm{if } k < T
\end{equation}
\begin{equation}
\label{eq2}
     P_k = (1-p)^{T} \hspace{1.45 in} \textrm{if } k = T
\end{equation}
\begin{equation}
\label{eq3}
    P_k =   \textcolor{black}{  p \cdot (1-p)^{T} \cdot (1-P_{\textrm{k-T-1}})}+ \textcolor{black}{P_{\textrm{k-1}}} \hspace{0.1 in}  \textrm{if } k > T
\end{equation}

\begin{figure}[!htb]
    \centering
        \vspace{-0.15 in}
\includegraphics[width= 2.5 in]{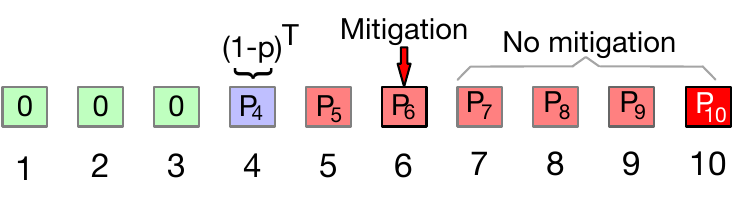}
    \vspace{-0.1 in}
    \caption{Sariou-Wolman model for getting $P_k$ for T=4. For failure at position-10,  position-6 must be failure-free, must get a mitigation, and none after.}
    \vspace{-0.1 in}
    \label{fig:swolman}
\end{figure}

Equation 1 and Equation 2 are trivial: (1) No failure with less than T activations and (2) at K=T activations, failure happens if all activations escape selection. For K$>$T activations, the recurrence is based on a powerful insight. For the row to fail exactly at the Kth activation, following must be true: (1) The position K-T must have received a mitigation  (hence the term \textcolor{black}{$p$}) (2) No mitigation since (hence the term \textcolor{black}{ $(1-p)^T$}) (3) Position K-T-1 must not already start with failure (hence the term \textcolor{black}{(1-$P_{\textrm{k-T-1}}$})). Finally, as failures  are cumulative, the term \textcolor{black}{
$P_{\textrm{k-1}}$}. Figure~\ref{fig:swolman} shows an overview of this model for T=4.

\subsection{Computing the Mean Time-to-Failure (MTTF)}

For a given number of activations to an attack row within tREFW, we use Equations 1-3 to estimate the failure probability ($P_{REFW}$) per tREFW. Based on Sariou-Wolman model~\cite{sariou}, we account for auto-refresh by reducing $P_{REFW}$ by a factor of (1-N/8192), where N denotes the length of the successful sequence in terms of tREFI. 
Equation~\ref{eq:ttbf} shows the {\em  Mean Time-to-Failure (MTTF)} for a bank. The MTTF for a system with $B$ banks would be  approximately $B$ times lower.
\begin{equation}
\label{eq:ttbf}
    \textrm{Mean-Time-to-Failure (Bank)} = \frac{1}{P_{REFW}} \cdot tREFW
    \vspace{0.1 in}
\end{equation}

\subsection{Minimum-Tolerated TRH: Key Figure-of-Merit}

We use a default {\em Target-MTTF} (per-bank) of 10,000 years (this is similar to the per-bank failure rate from naturally occurring errors~\cite{ddr4errors}, sensitivity in Section~\ref{sec:mttf}).  We define the {\em Minimum Tolerated TRH} ({\bf MinTRH}) as the {\em lowest TRH} for which the design can meet the Target-MTTF.  We denote {\bf MinTRH-D} as the per-row TRH for a double-sided pattern.

\newpage
\ignore{
TODO: Ideally, we need only a single entry

}

\newpage
\section{Minimalist In-DRAM Tracker (MINT)}

Secure trackers require significant storage, whereas existing low-cost trackers have been rendered insecure.  To better understand this dichotomy, and develop a low-cost and secure tracker, we classify tracker design-space into three types, depending on how they select the row to be mitigated at a given REF. First, {\em past-centric}, which considers past behavior, for example, selecting the row with the highest counter value. To do a reliable selection, significant amount of past behavior is needed.  Second, {\em present-centric}, which makes selection decision purely based on the currently activated row (say select the given row with a given probability and store it in a single-entry tracker).  Unfortunately, such {\em \textcolor{black}{InDRAM-PARA}}, suffers from the problems of non-uniform mitigation probability over tREFI and non-selection (even if the row is continuously activated in the tREFI interval). Therefore, it has a high MinTRH. We propose a {\em Minimalist In-DRAM Tracker (MINT)}, which is {\em future-centric} and provides a secure mitigation with just a single-entry.  We first provide an overview and design of MINT, then derive the security for worst-case pattern, then discuss the impact of {\em Transitive Attacks} and {\em Spatial Attacks}.

\subsection{Overview of MINT: A Single-Entry Tracker}

\begin{figure}[!htb]
    \centering

\includegraphics[width= 3.4 in ]{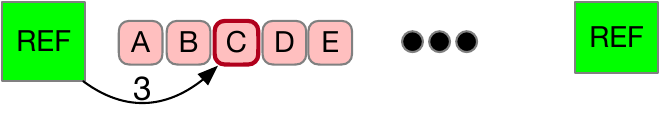}
   
    \caption{Overview of MINT. At each REF, MINT decides (a-priori) which activation number will get selected for mitigation at next REF.}
  
    \label{fig:overview}
\end{figure}

Figure~\ref{fig:overview} shows an overview of MINT. Lets say each tREFI window can have up-to $M$ activations.  At each REF, MINT decides which of the {\em future} M activations must be selected for mitigation at the next REF. It uses a {\em Uniform Random (URAND)} selection for all possible M positions. Each activation within tREFI is given a sequence number and when it reaches the selected number, the given row (e.g. Row-C), is selected for getting mitigated at the next REF. The process repeats at each subsequent REF with a new URAND selection. 

Note that the selection of MINT is done without knowing which address will appear at the selected activation number.  By design, MINT does not suffer from overwrite problem of IID-Tracking (as no more than one row can be selected for mitigation). Furthermore, if a given address occurs $M$ times in the window, then it is guaranteed to be selected by MINT.


Note that, if an access pattern has fewer than $M$ activations within tREFI, then MINT can  sometimes not select any row for mitigation at the next REF. However, this does not impact the security of MINT as the slot not used for activation can be treated, for the purpose of analyzing security, as an activation to a decoy (benign) row. For the attacker to have the highest chance of success, an attack pattern must use all the activation slots for causing {\em damage}. Therefore, for our security analysis, it is safe to assume that all M activations are used in an attack. 


\subsection{Design and Operation of MINT}

\begin{figure}[!htb]
\vspace{-0.15 in}
    \centering
\includegraphics[width= 3.2 in ]{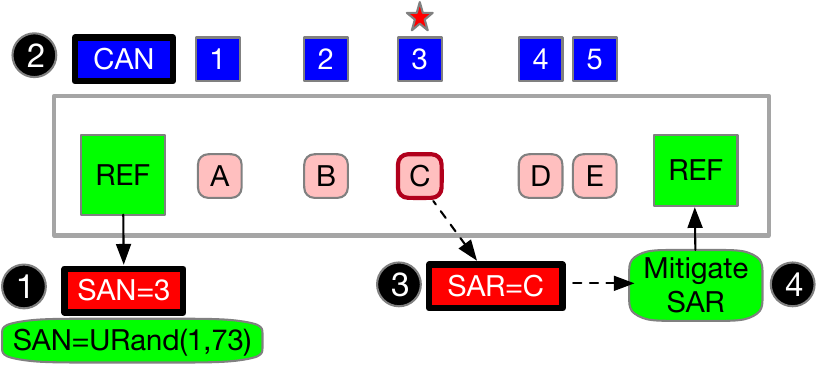}
    \caption{Design and Operation of MINT. At each REF, SAN is set with URAND. During tREFI, CAN tracks sequence number for ACT. If CAN=SAN, the row address is stored in SAR. At REF, SAR is mitigated. }
    \vspace{-0.15 in}
    \label{fig:mint}
\end{figure}

Figure~\ref{fig:mint} shows the design and operation of MINT. MINT consists of three registers: (1) {\em Selected Activation Number (SAN)}, which stores the activation number that will be selected in the upcoming interval (2) {\em Current Activation Number (CAN)}, which provides a sequence number to each activation in the tREFI window, and (3) {\em Selected Address Register (SAR)}, which stores the address of the row to be mitigated at the next REF. SAR contains a valid bit to indicate if the SAR is filled.

The maximum number of activations within the tREFI window is 73 (so, M=73).  \circled{1} At each REF, the SAN is initialized based on a URAND selection of all 1 to 73 slots. The CAN is reset to 0. The SAR is set to invalid. \circled{2} During the tREFI, the CAN provides a sequence number to each activation. For example, the first activation has CAN=1, second has CAN=2, and so on. \circled{3} When CAN is equal to SAN, it indicates that the given address is selected for mitigation, which means the row address is copied to SAR and SAR is set to be valid. \circled{4} At the next REF, if the SAR is valid, the row-address stored in SAR gets mitigated.  For mitigation, we assume that the number of victim rows equivalent to the {\em Blast Radius} is refreshed on either side of the given aggressor row.  The process repeats at each subsequent REF. MINT is a {\em single-entry tracker}, as only the {\em Selected Address Register (SAR)} holds the address of the row to be mitigated. 

\subsection{Impact of Classic Attacks: The Need for New Patterns}

By design, MINT is robust against classic single-sided and double-side patterns that operate continuously during tREFI.  

\vspace{0.05 in}
\noindent{\bf Single-Sided Attack:} If a pattern repeatedly activates a given row, say Row-A (we use closed-page policy), during the entire tREFI window, then MINT is guaranteed to select this row for mitigation. Thus, MINT would limit such a classic single-sided attack to at-most $M$ activations on the attacked row. 

\vspace{0.05 in}
\noindent{\bf Double-Sided Attack:} If a pattern repeatedly activates a pair of rows in alternating fashion and the pair shares a victim, say Row-A and Row-C with shared victim Row-B, then MINT is guaranteed to select one of the two rows (Row-A or Row-C), which will cause a refresh of victim (Row-B).  Thus, MINT would limit the effect on victim to at-most M activations. 
\vspace{0.05 in}

As classic attack patterns are not useful for analyzing the security of MINT, we investigate the worst-case pattern for MINT and use it to determine the MinTRH.

\subsection{Estimating the MinTRH of MINT}

Our analysis is based on three observations for MINT: (1) The selection decisions are localized to within the tREFI, so what happens in the previous or next tREFI does not impact the selection during current tREFI. (2) The probability of a row getting selected does not depend on the position in the tREFI interval, as all positions are equally likely to get selected, therefore reordering the address pattern within tREFI does not impact security  (3) If a given row is activated $n$ times within the same tREFI, it is $n$ times more likely to get selected for mitigation (in the limit, if n=73, the row has guaranteed selection). So, successful attacks must avoid having a large number of activations to the same row within the single tREFI.  These properties can help us identify the worst-case pattern.

\vspace{0.05 in}
\noindent{\bf Pattern-1: Single-Row, Single-Copy:} This pattern focuses the attack on a single row (Row-A). During each tREFI, it performs only a single activation on Row-A and the remaining 72 activation slots remain unused. The pattern repeats 8192 times. During each tREFI, each activation of Row-A will get selected with probability p=1/73. We use the Sariou-Wolman model to determine the probability of failure ($P_{REFW}$) on the 8192nd activation. We use the $P_{REFW}$ to determine MinTRH.  For this pattern, the {\bf MinTRH is 2461}.

\begin{figure}[!htb]
\vspace{-0.05 in}
    \centering
\includegraphics[width= 3.5 in ]{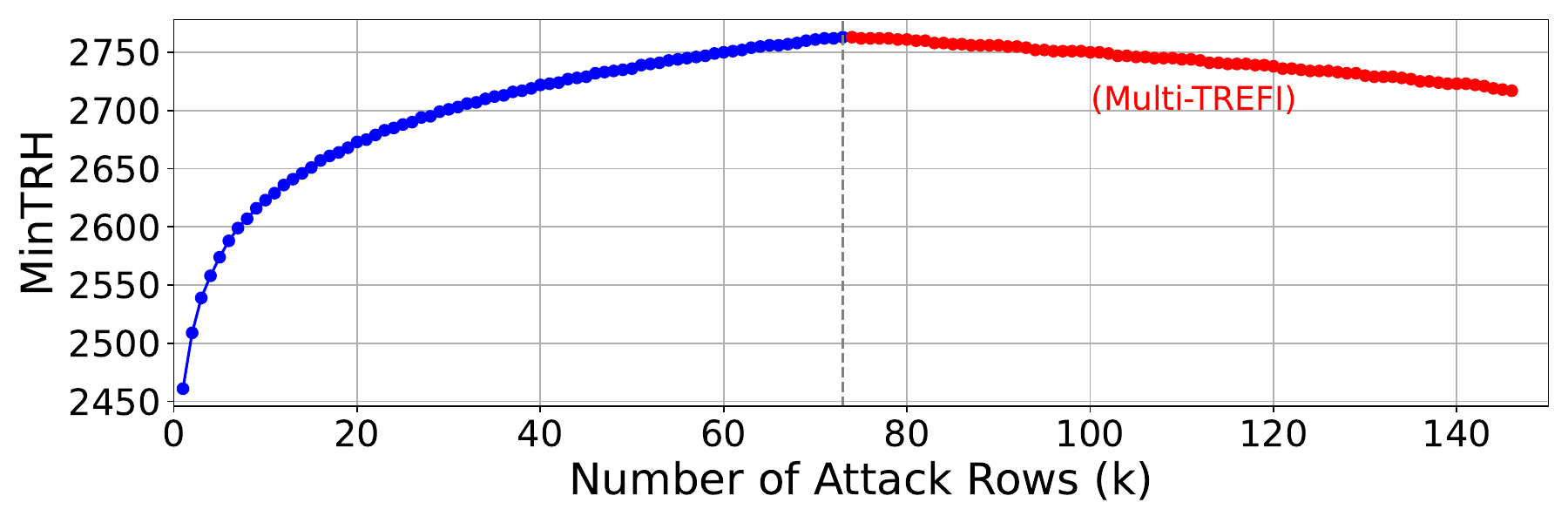}
    \caption{MinTRH for pattern-2 as the number of attack lines (k) is varied. }
    \vspace{-0.1 in}
    \label{fig:circ}
\end{figure}

\vspace{0.05 in}
\noindent{\bf Pattern-2: Multi-Row, Single-Copy:} For faster attacks, the pattern must try to use all of the 73 slots in the tREFI interval.  A simple way to achieve this is to perform a single activation to $k$ attack rows within the single tREFI. If $P_{REFW}(1)$ is the failure probability for pattern-1, then with $k$ lines, the failure probability increases by $k$ times.  So, $P_{REFW}(k) = k \cdot P_{REFW}(1)$. We use $P_{REFW}(k)$ to determine the MinTRH for each k.  Figure~\ref{fig:circ} shows the MinTRH as $k$ is varied from 1 to 73. We also evaluate a {\em multi-TREFI} attack contains more than 73 rows. The MinTRH increases with k, peaks at k=73, and reduces thereafter for multi-TREFI pattern, as activations per row reduces. For pattern, with k=73, the {\bf MinTRH is 2763}.

\begin{figure}[!htb]
\vspace{-0.05 in}
    \centering
\includegraphics[width= 3.5 in ]{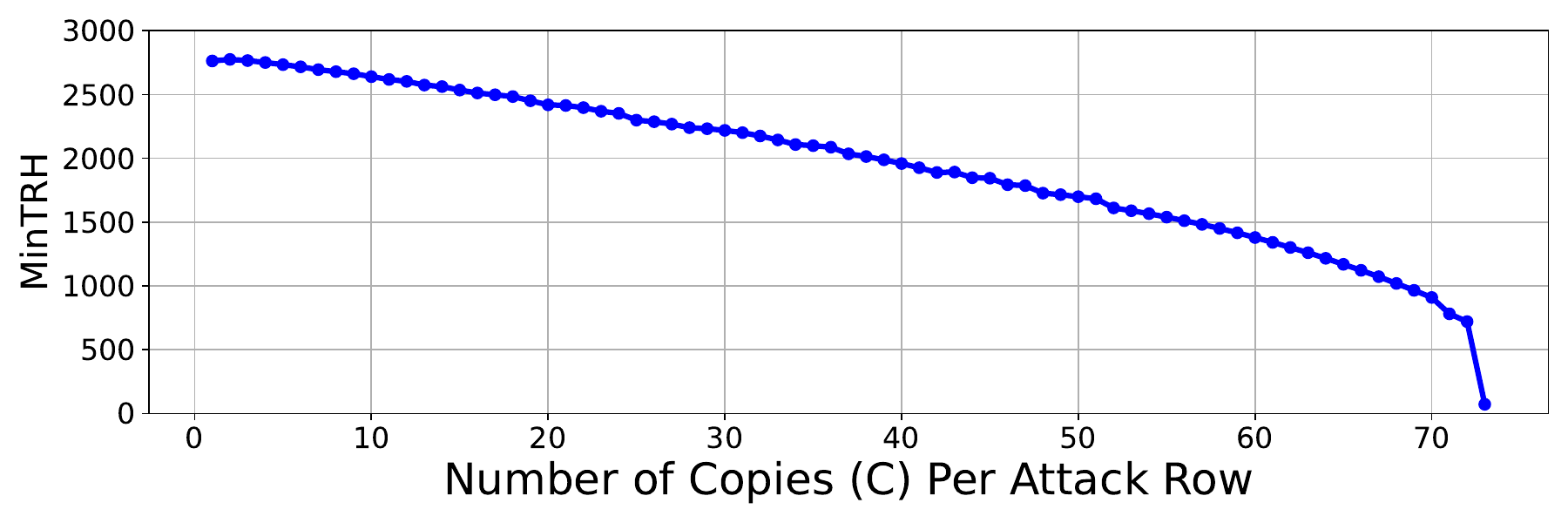}
    \caption{MinTRH for pattern-3 as the number of times a given attack row is activated within the pattern (copies) gets varied from 1 to 73. }
    \vspace{-0.2 in}
    \label{fig:copies}
\end{figure}

\vspace{0.05 in}
\noindent{\bf Pattern-3: Multi-Row, Multi-Copy:} An attacker could spend the 73 activations on attacking $k$ rows but activate each row $c$ times. This attack is less stealthy as each row has $c$ times higher chance of getting selected in each tREFI.  Figure~\ref{fig:copies} shows the MinTRH of MINT as the number of copies (c) per row is varied from 1 to 73 (the number of rows is adjusted to fit in one tREFI). With few copies (1-3), the MinTRH of pattern-3 remains similar (within 0.5\%) to pattern-2, however, it drops significantly for 4+ copies.  Thus,  having a large number of copies within tREFI is not an effective attack.

\noindent{\bf Key Takeaway:} To develop the most effective {\em direct} attacks for MINT, the attacker is forced to have only 1 activation to any attack row within tREFI to maximize stealth. Under this constraint,  we can  use our analysis with pattern-2 (with 73 attack rows) to determine that MINT has {\bf MinTRH of 2763}.

\subsection{Impact of Transitive Attacks}
\label{sec:transitive_attacks}

\begin{figure}[!htb]
    \centering
\includegraphics[width= 3.2 in]{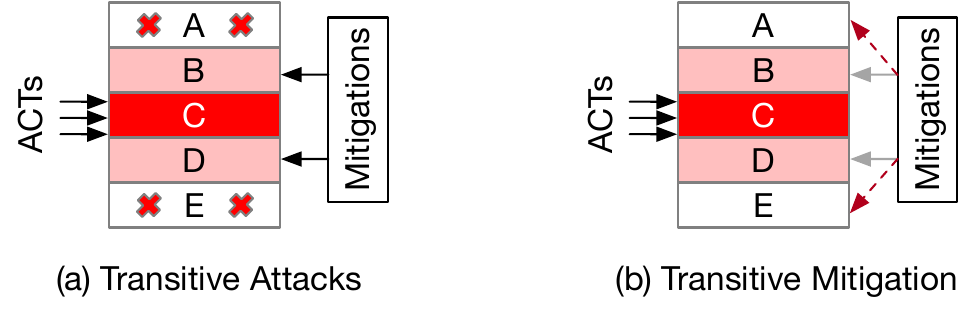}
    \vspace{-0.1 in}
    \caption{(a) Transitive Attacks (B) Extending MINT with Transitive Mitigation, which  performs mitigative refreshes on the victim of victim-rows.}
    \label{fig:transitive}
\end{figure}

Thus far, we have only been focused on {\em direct} attacks, which aim to directly cause failures in the victim rows of a given row.  {\em Transitive Attacks}~\cite{DRAMSecrecy}, such as Half-Double~\cite{HalfDouble}, offer another indirect way to cause failure. Figure~\ref{fig:transitive} (a) shows an example of such an attack, where Row-C receives continuous activations (using a single-sided attack).  MINT will mitigate the neighbors of Row-C at each REF, causing 8192 victim refreshes on Row-B and Row-D. The activations from these mitigative refreshes are silent (not observable by MINT) and can cause failures in Row-A and Row-E.  Thus, such an attack increases MinTRH of MINT to 8192. Note that refreshing two rows on either side of an aggressor does not mitigate transitive attacks, as the third row now experiences failures. 


We extend MINT to be secure against transitive attacks by using {\em transitive mitigation}. A transitive mitigation refreshes the victim of victim-rows.  For example, a regular mitigation for Row-C would refresh Row-B and Row-D, however, a transitive mitigation would refresh Row-A and Row-E. Ideally, on each activation, if we do normal mitigations with probability p, we should do transitive mitigations with probability $p^2$. 

Instead of choosing from 73 slots, we modify MINT to select from 74 slots (the extra slot indicating transitive mitigation for the recently mitigated row).  As there are 74 slots, now URAND must be modified to select (0-73) positions, where 0 indicates transitive mitigation for the current address in SAR (if SAN is 0, SAR is preserved at REF, and indicates transitive mitigation). We note that the transitive mitigation can be applied recursively~\cite{PRIDE} (distance is increased if SAN=0 comes consecutively).  As the probability of selection is reduced to 1/74, per pattern-2, the {\bf MinTRH of MINT is 2800.}

\newpage

\subsection{Impact of Spatial-Correlation Attacks}

\begin{figure}[!htb]
    \centering
\includegraphics[width= 3.4 in]{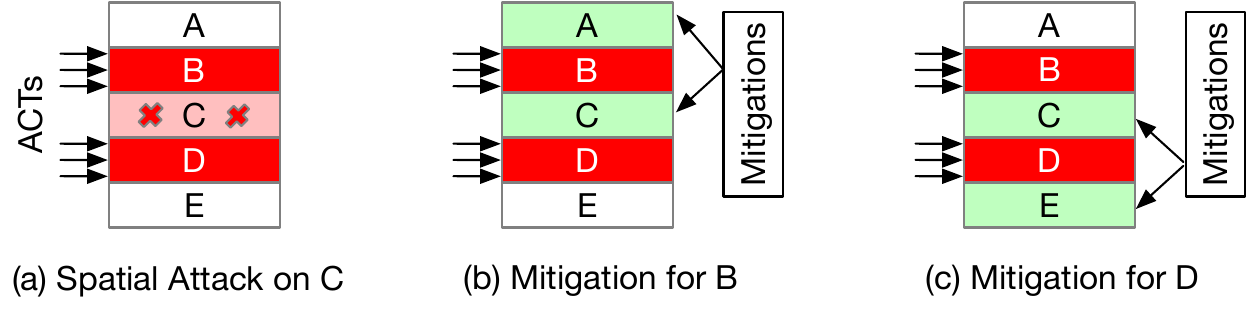}
    \caption{Sptail correlation attacks (a) Double-sided attack on Row-C (b) Mitigation for B refreshes A and C (c) Mitigation for D refreshes C and E. 
 Thus, C has as many chances of mitigations as summed over B and D. }
    \label{fig:spatial}
\end{figure}

Our analysis thus far has assumed that the rows in a multi-row attack (pattern-2) are not spatially correlated.  However, an attacker could try to sandwich a victim-row, between two attack rows to increase the hammers suffered by the victim row.  This type of spatial correlation attack is present in the double-sided pattern, as shown in Figure~\ref{fig:spatial}(a), where victim Row-C is between two aggressor rows, Row-B and Row-D. 

Such a pattern is challenging for prior counter-based trackers, that determine the selection decision based on counter-values. If both aggressor rows perform T activations before either one gets selected for a mitigation, then the victim-row is subjected to  2T activations. Thus, the effective threshold tolerated by  counter-based schemes get doubled due to such a pattern. MINT is immune against such spatial patterns.

MINT performs probabilistic selection.  If Row-B is selected, then Row-A and Row-C get refreshed, as shown in Figure~\ref{fig:spatial}(b).  If Row-D is selected, then Row-C and Row-E get refreshed, as shown in Figure~\ref{fig:spatial}(c).  Thus, Row-C has as many chances for refresh as offered by {\em both} Row-B and Row-D. For example, if Row-B and Row-D each had 2800 activations (MinTRH), then Row-C gets 5600 chances of refresh.   

The MinTRH of MINT indicates how many chances of mitigation can be escaped successfully.  Thus, the total number of activations over Row-B and Row-D cannot exceed 2800, so, on average, each of the two rows can have up-to 1400 activations.  This analysis also helps us bound the {\em Minimum Tolerated TRH (MinTRH-D)} for a double-sided pattern. Note that, device characterization studies typically report TRH in terms of per-row activations for a double-sided pattern, so they report MinTRH-D.   The {\bf MinTRH-D of MINT is 1400}.

\subsection{Comparison with Prior Trackers}

We compare MINT with four designs, two counter-based (PRCT and Mithril~\cite{kim2022mithril}) and two probabilistic (PARFM~\cite{kim2022mithril} and \textcolor{black}{InDRAM-PARA}).  \textcolor{black}{InDRAM-PARA} is our adaptation of PARA~\cite{kim2014flipping} to the in-DRAM setting, using a single-entry tracker, in order to highlight the differences between MINT and an alternative probabilistic selection.  We compare these designs in term of tolerated TRH (MinTRH-D), the number of tracking entries, and the impact of Transitive Attacks.  

\vspace{0.05 in}
\noindent{\bf PRCT:}  This is an idealized past-centric scheme that maintains a counter for each row.  As all activations (including from victim refresh) increment the counter, PRCT is immune to transitive attacks. To determine MinTRH of PRCT, we use the {\em Feinting Attack}~\cite{ProTRR}, and start by spreading the available activations across 8192 aggressor rows. At each REF, the row with the highest counter value gets mitigated\footnote{This design always picks a row to be mitigated as long as there is at-least one activation within the tREFI.  The energy overheads of mitigation for such a design may be high (in the worst-case, one mitigation per ACT), so a practical PRCT design is likely to mitigate only if the maximum counter value is above some threshold.  We do not put such constraints on our PRCT.} and removed from the list of aggressor rows. In the second-to-last round (8191), all the activations within tREFI are focused on the last two remaining aggressor rows. To determine, MinTRH-D, we assume that the victim is placed in between these two aggressor rows.  The MinTRH-D of PRCT is 623.

\vspace{0.05 in}
\noindent{\bf Mithril:}  This counter-based past-centric design tracks only a subset of rows, and uses a proactive mitigation (i.e. it selects the row with the highest counter value for mitigation at each REF).  As activations from mitigative refreshes increment the counter, this design is immune to transitive attacks.  We use the Theorem-1 provided in the original paper~\cite{kim2022mithril} to derive the number of counters required for a particular MinTRH-D. Mithril requires at-least 677 entries to get MinTRH-D of 1400.

\vspace{0.05 in}
\noindent{\bf PARFM:}  This is a past-centric probabilistic design that buffers all the activations during the tREFI window, and on reaching REF, it randomly selects one of the buffered entries to get mitigated, and invalidates all the buffered entries to free up space for the next tREFI.  As there could be up-to 73 activations within the tREFI interval, this design requires an overhead of 73 entries per bank. As only the demand activations are used for selection, this design is vulnerable to transitive attacks.  The MinTRH-D of PARFM is 4096. 

\vspace{0.05 in}
\noindent{\bf \textcolor{black}{InDRAM-PARA}:}  This is a current-centric probabilistic design.  On each activation, a row is selected with probability {\em p} (p=1/73 for our study) and stored in a single-entry tracker. This row is mitigated only if it can {\em survive} the remaining time in tREFI, as other selections can overwrite this entry.  We derive the MinTRH-D of the \textcolor{black}{InDRAM-PARA} to be 3732. As the MinTRH of this design is relatively high, direct attacks allow more unmitigated activations on a given row than transitive attacks, therefore, this design is immune to transitive attacks.

\vspace{0.05 in}

Table~\ref{tab:trackers} compares the four trackers with MINT in terms of type, MinTRH-D, entries, and vulnerability to Transitive Attacks.  
MINT has a MinTRH similar to a 677-entry Mithril tracker, and has 2.25x the MinTRH of the idealized PRCT design. This bound with PRCT becomes within 2x under refresh postponement, which we discuss next.

\begin{table}[htb]
  \centering
  \caption{Comparison of in-DRAM Trackers }
  \label{tab:trackers}
  \begin{tabular}{ccccc}
    \hline
   \textbf{Design} & \textbf{Type} & \textbf{MinTRH-D} & \textbf{Entries} & \textbf{Transitive}\\ 
                    & \textbf{(Centric)}  & \textbf{(Threshold)} & \textbf{(Per-Bank)} & \textbf{Attacks}\\ \hline 
   PRCT    & Past  & 623   & 128K & Immune \\ 
   Mithril    & Past  & 1400   & 677 & Immune \\ 
   PARFM           & Past  & 4096   & 73 & Vulnerable \\  
   \textcolor{black}{InDRAM-PARA}       & Current  & 3732   & 1 & Immune \\ 
   MINT           & Future  & 1400   & 1 & Immune \\ \hline  
  \end{tabular}

\end{table}

\newpage
\section{Handling Refresh Postponement}

\begin{figure}[!htb]
    \centering
       \vspace{-0.05 in}
\includegraphics[width= 2.5 in]{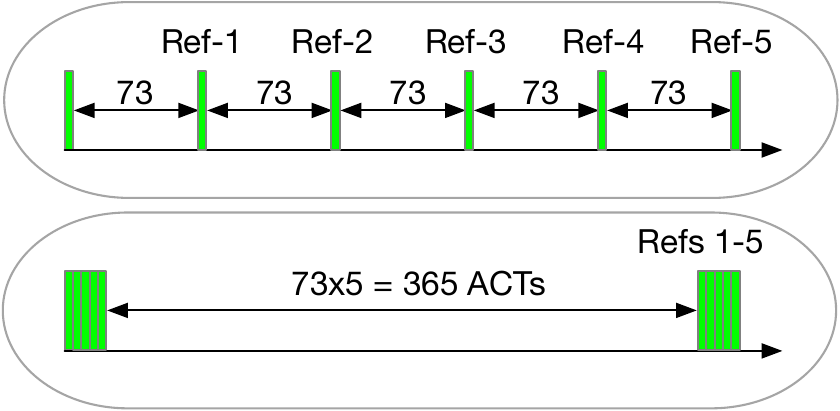}
    \vspace{-0.05 in}
    \caption{Refresh postponement in DDR5 (top) Timely refresh (bottom) Batches of 5 refresh allowing up-to 365 ACTs between refresh.}
    \vspace{-0.1 in}
    \label{fig:refdelays}
\end{figure}

Thus far, we assumed that a refresh is performed at every tREFI.  However, in reality, DDR5 specifications allow the postponement of up-to 4 refresh operations. A maximum of 5 refreshes can be batched and performed together, as shown in Figure~\ref{fig:refdelays}. Refresh postponement increases the number of activations between refresh from 73 to 365.

\subsection{Impact of Refresh Postponement on Optimal Trackers}

Refresh postponement can allow the attacker to cause more activations on the aggressor row (selected for mitigation) during the period of refresh postponement.  The impact of refresh postponement is easier to understand for counter-based trackers~\cite{ProTRR}, as the threshold gets revised by an amount equal to the additional ACTs due to refresh postponement (so 73x4 = 292 in our case, split equally on either side of a double-sided attack).  Thus, the MinTRH-D of PRCT increases from 623 to 769. Similarly, Mithril requires number of entries (per-bank) to increase from 677 to 827 to maintain MinTRH-D of 1400.

\subsection{Impact of Refresh Postponement on Low-Cost Trackers}

To the best of our knowledge, no prior work has studied the impact of refresh postponement on low-cost trackers.  Refresh postponement is especially problematic for low-cost trackers as they track only a few entries, and are tailored for a given rate of mitigation (e.g. one per tREFI).

If the maximum number of activations within tREFI is M (73 in our case), then refresh postponement can make all activations past M {\em invisible} to the tracking mechanism. This is the case for both MINT and PARFM, as they are designed for only M  activations within tREFI and a refresh thereafter.  With refresh postponement, the attacker can perform activations on decoy rows during the first M activations, and then deterministically perform 4M activations on the attack row (without receiving any mitigations). Thus, refresh postponement can allow the attacker to deterministically perform {\bf 478K} (!) activations on a row every 32ms using MINT and PARFM.

For the \textcolor{black}{InDRAM-PARA}, the extra activations during the period of refresh postponement can cause the tracked entry to get dislodged easily. If the attacker places the attack row in the first position, then the probability of survival after another 364 activations is 0.66\%. Refresh postponement increases the  MinTRH-D of the IID tracker from 3.7K to more than  {\bf 21K.}

Thus, refresh postponement demolishes existing low-cost trackers. We propose a generalized design that makes low-cost trackers compatible with refresh postponement.

\subsection{Practical Solution: Delayed Mitigation Queue (DMQ)}

As refresh postponement is in DDR5 standards, all trackers must support refresh postponement. To achieve this, we propose a general solution, {\em Delayed-Mitigation Queue (DMQ)}.

\begin{figure}[!htb]
    \centering
       \vspace{-0.05 in}
\includegraphics[width= 2.4 in]{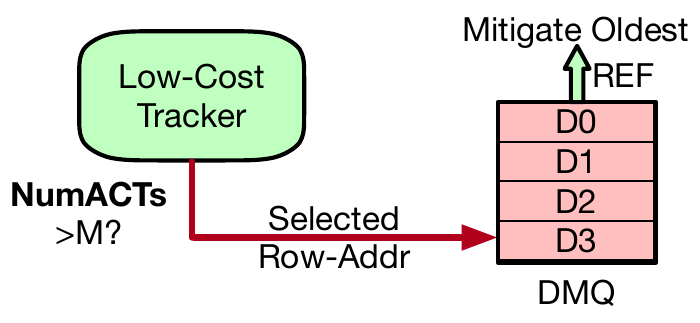}
    \vspace{-0.05 in}
    \caption{Design of Delayed-Mitigation Queue (DMQ). DMQ makes  low-cost in-DRAM trackers compliant with refresh postponement.}
    \vspace{-0.05 in}
    \label{fig:dmq}
\end{figure}

Figure~\ref{fig:dmq} shows the overview of our design.  We consider a generic low-cost tracker (e.g. MINT, PARFM, \textcolor{black}{InDRAM-PARA} etc.).  DMQ requires the tracker to count the number of activations {\em (NumACTs)} since the last refresh (MINT already does this with CAN, but other trackers may need an additional register).  If NumACTs exceeds the maximum number of activations in tREFI (e.g. 73), it is reset to 1, and the tracker perform a {\em pseduo-mitigation}. During pseudo-mitigation,  the tracker provides the address of the selected aggressor row (e.g. stored in SAR for MINT and \textcolor{black}{InDRAM-PARA}), which is inserted into a FIFO buffer called the {\em DMQ}. The DMQ has four entries (as up-to four refreshes can be postponed). On REF, if the DMQ contains at-least one valid entry, the oldest entry from the DMQ is mitigated. Else, the tracker selects and mitigates normally, similar to no refresh postponement.

\subsection{Bounding the Impact of DMQ on Tolerable TRH}


As DMQ is a FIFO, the maximum number of activations for which the row selected by the tracker will get delayed in receiving a mitigation while waiting in the DMQ is 292 (73x4). At the worst-case, the pattern may be accessing the same row continuously, in which case the row may receive up-to 292 activations while in the DMQ, so the MinTRH of the tracker would increase by 292, or equivalently, the MinTRH-D would increase by 146.  Thus, DMQ makes the impact of refresh postponement on low-cost trackers, similar to counter-based trackers. Table~\ref{tab:dmq} shows the impact of refresh postponement (with and without DMQ). As MINT forces a pattern that has a single activation of a row within tREFI, delaying the mitigation by four tREFI causes only four more activations, so the MinTRH-D of MINT increases by 4 to 1404 (it can be increased to 1482* using an adaptive attack (details in Appendix B). {\bf The MinTRH-D of MINT is  1482 (outperforming 677-entry Mitrhil and 1.9x of PRCT}).


\begin{table}[htb]
  \centering
    \vspace{-0.1in}
  \caption{Impact of Refresh Postponement and DMQ on Trackers }
  \label{tab:dmq}
  \begin{tabular}{ccccc}
    \hline
   \textbf{Design} & \textbf{Entries} & \textbf{MinTRH-D} & \textbf{MinTRH-D} & \textbf{MinTRH-D}\\ 
                    & \textbf{(Bank)}  & \textbf{(NoPostpone)} & \textbf{(No DMQ)} & \textbf{(with DMQ)}\\ \hline 

   PRCT             & 128K  & 623   & 769 & 769 \\ 
   Mithril          & 677 & 1400   & 1546 & 1546 \\ 
   PARFM            & 73  & 4096   & 478K & 4242 \\  
   \textcolor{black}{InDRAM-PARA}      & 1  & 3732   & 21.3K & 3650 \\ 
   MINT             & 1  & 1400   & 478K & 1404/1482* \\ \hline  
  \end{tabular}
  \vspace{-0.15 in}

\end{table}

\begin{figure*}[!tbh]
    \centering
    \includegraphics[width=6.8in]{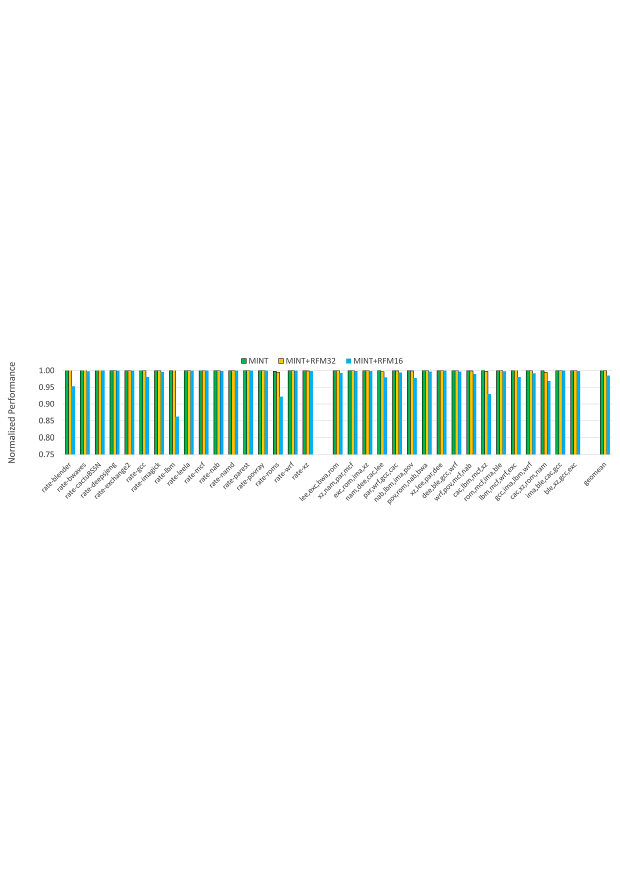}
    \vspace{-0.15 in}
    \caption{Normalized performance. MINT incurs zero slowdown. MINT+RFM32 and MINT+RFM16 incur 0.2\% and 1.6\% slowdown, respectively.}
    \label{fig:perf}
   \vspace{-0.15 in}
\end{figure*}

\section{Scaling to Lower Thresholds with RFM}
\label{sec:RFM}

The MinTRH of MINT can be reduced by increasing the mitigation-rate. DDR5 specifications include {\em Refresh Management (RFM)} that enables the memory controller to provide more time to the DRAM chip to perform mitigation.  
RFM enables the memory controller to issue additional mitigations (RFM commands) to the DRAM when the activations per bank crosses a threshold. 
However, the actual row to be mitigated still depends on the in-DRAM tracker. The memory controller maintains a {\em Rolling Accumulation of ACTs (RAA)} counter ~\cite{kim2022mithril}) per bank. When any RAA counter crosses a threshold, $RFM_{TH}$, the memory controller resets the RAA counter and issues an RFM to the corresponding DRAM bank.

We co-design MINT with RFM to scale to lower thresholds.  We evaluate two variations: with a RFMTH of 32 (MINT+RFM32) and 16 (MINT+RFM16). As the number of activations within the mitigation period is now 32 or 16, we modify MINT to select URAND(0,32) or URAND(0,16).

Table~\ref{tab:rfm} shows the MinTRH-D for MINT and MINT+RFM. We evaluate various mitigation rates, including 0.5x (one mitigation every two tREFI), 1x (one mitigation every tREFI), RFM32 ($\sim$2x mitigation rate), and RFM16 ($\sim$4x mitigation rate). As RFM can also be delayed by 3x-6x, we implement MINT+RFM with DMQ and report the threshold under an adaptive attack. MINT+RFM16 has a {\bf MinTRH-D of 356}.

\begin{table}[htb]%
  \centering
  \caption{The MinTRH-D of MINT and MINT+RFM (includes DMQ)}
  \label{tab:rfm}
  \begin{tabular}{ccc}
    \hline
   \textbf{Scheme} & \textbf{Relative Mitigation Rate} & \textbf{MinTRH-D}\\ \hline 
    MINT          & 0.5x (one per two tREFI)        & 2.70K\\ 
    MINT          & 1x (one per  tREFI)             & 1.48K \\ 
    MINT+RFM32    & 2x (approx two per tREFI)      & 689\\ 
    MINT+RFM16    & 4x (approx four per tREFI)     & {\bf 356} \\ \hline 
  \end{tabular}
\end{table}

\section{Results}
\subsection{Impact on Performance} 

We model MINT and MINT+RFM in Gem5~\cite{lowe2020gem5} simulator. Table~\ref{table:system_config} shows our configuration. We model a 4-core out-of-order CPU with DDR5, using Micron Datasheet~\cite{micron_ddr5}. We evaluate our design with 17 SPEC2017~\cite{SPEC2017} \textit{rate} workloads and 17 mixed workloads. We skip 25 billion instructions and simulate 250 million instructions. Per DDR5 specifications, we assume tDRFM$_{sb}$ delay equal to tRFC (410ns) while tRFM$_{sb}$ delay equal to half the time of tRFC (205ns). We assume that there is no rate-limit to the number of DRFM commands issued per tREFI (JEDEC limits one DRFM per two tREFI).

\begin {table}[htb]
\begin{footnotesize}
\begin{center} 
\caption{Baseline System Configuration}
\begin{tabular}{|c|c|}
\hline
  Out-of-Order Cores           & 4 core, 3GHz, 8-wide, 192-ROB   \\
  Last Level Cache (Shared)    & 4MB, 16-Way, 64B lines \\ \hline
  Memory specs                 & 32 GB, DDR5 \\
  t$_{RCD}$-t$_{CL}$-t$_{RP}$-t$_{RC}$ & 16-16-16-48 ns\\
  Banks x Ranks x Channels     & 32$\times$1$\times$1 \\
  Rows                & 128K rows, 8KB row buffer\\ \hline
\end{tabular}
\label{table:system_config}
\end{center}
\end{footnotesize}
\end{table}


Figures~\ref{fig:perf} shows the relative performance of MINT and MINT+RFM, normalized to the DDR5 baseline.  MINT incurs zero slowdown, as the mitigations required for MINT are performed within the tRFC period as mentioned in DDR5 specifications. MINT+RFM32 incurs negligible slowdown (0.1\%) and MINT+RFM16 incurs an average slowdown of  1.6\%. Thus, MINT and MINT+RFM provide a scalable and low overhead mitigation for Rowhammer, even at low thresholds.

\subsection{Impact of Target Time-to-Fail on Threshold}
\label{sec:mttf}
As MINT is a probabilistic design; we deem it to be secure if the time-to-fail is greater than the {\em Target Time-to-Fail (Target-TTF)}.  We used a default Target-TTF of 10,000 years per bank as it is similar to per-bank failure-rate similar from naturally occurring errors~\cite{ddr4errors}. Table~\ref{tab:mttf} shows MinTRH-D of MINT for varying Target-TTF (per-bank) and MTTF for our system (64 banks, but only 22 can be used concurrently due to tFAW). MINT provides several decades/centuries of protection even under continuous attacks on all available banks.

\begin{table}[htb]
  \centering
    \vspace{-0.1in}
  \caption{The MinTRH and MinTRH-D of MINT for Various Target-TTF  }
  \label{tab:mttf}
  \begin{tabular}{ccccc}
    \hline
   \textbf{Target-TTF } & \textbf{MTTF } & \textbf{MinTRH-D} & \textbf{MinTRH-D} & \textbf{MinTRH-D}\\ 
    \textbf{(Bank) } & \textbf{(System)} & \bld{MINT} & \bld{ (+RFM32)} & \bld{ (+RFM16)}\\ \hline
    1K years &  45 years  & 1.40K   &   651 &  336 \\  
    \bld{10K years} &  \bld{450 years}  & \bld{1.48K}   &   \bld{689} &  \bld{356} \\  
      100K years &  4.5K years  & 1.57K   &   726 &  375 \\  
     1Million years &  45K years  & 1.64K   &   763 &  395 \\  \hline
  \end{tabular}

\end{table}

\subsection{Storage Overheads}

MINT requires a CAN (7-bits), SAN (7-bits) and SAR (18 bits), for a total of 32 bits (4 bytes).  The DMQ requires 4 entries (19 bits, including one for transitive-mitgation) for a total of 9.5 bytes.  Thus, MINT+DMQ requires less than 15 bytes per bank. 
It also requires an in-DRAM pseudo random number generator, similar to DSAC~\cite{DSAC} and PAT~\cite{HynixRH}.




\subsection{Energy Overheads}

\color{black}
The energy overheads of MINT can be attributed to three sources (1) the random number generator (RNG) that is consulted at each tREFI to derive the SAN, (2) the additional structure of DMQ, and (3) the extra activations to perform mitigative refreshes. Our energy overheads include the energy incurred in all three sources. 

MINT uses a 7-bit TRNG ~\cite{katz2008robust, yu2019survey} which consumes 90 micro-watts of static power and 200 micro-watts of dynamic power. The total power of the TRNG (290 micro-watts) is 4 orders of magnitude lower than the DRAM power.

We use CACTI-6.5 to estimate the DMQ power. DMQ consumes static power of 48 micro-watts and dynamic power of 38 micro-watts. The total power of DMQ (86 microwatts) is 4 orders of magnitude lower than the DRAM power. 

Table~\ref{table:energy} shows the relative memory energy consumption (including DMQ and RNG) of MINT and MINT+RFM normalized to the baseline.  To determine DRAM energy, we use Gem5 and memory-power model. Overall, MINT incurs only 1\%-3\% extra energy over an idealized baseline that incurs no energy overheads for mitigation.  MINT increases ACT energy by 6\%--25\%, however, as activation energy is only 13\% of the total energy, its impact on overall energy is small.

\begin{table}[!htb]
  \centering
  \vspace{-0.1in}
  \caption{Memory Energy Overheads of MINT and MINT+RFM}
  \vspace{-0.1in}
  \begin{footnotesize}
  \label{table:energy}
  \begin{tabular}{cccc}
    \hline
    \textbf{Config} & \textbf{\new{ACT Energy}} & \textbf{\new{Non-ACT Energy}} & \textbf{\new{Total }}  \\ \hline \hline
    \rule{0pt}{0.8\normalbaselineskip}
    Base (No Mitig) & \new{1x (13\% overall)} & \new{1x (87\% overall)}  & \new{1x} \\ \hline
    MINT           & \new{1.06x} & \new{1.00x}  & \new{1.01x} \\ 
    MINT+RFM32     & \new{1.10x} & \new{1.00x}  & \new{1.01x} \\ 
    MINT+RFM16     & \new{1.25x} & \new{1.01x}  & \new{1.03x} \\ \hline 
  \end{tabular}
  \end{footnotesize}
\end{table}

\color{black}

\ignore{
\begin{table}[!htb]
  \centering
  \vspace{-0.1in}
  \caption{DRAM Energy Overheads of MINT and MINT+RFM}
  \vspace{-0.1in}
  \begin{footnotesize}
  \label{table:energy}
  \begin{tabular}{cccc}
    \hline
    \textbf{Config} & \textbf{Static Energy} & \textbf{Dynamic Energy} & \textbf{Total Energy}  \\ \hline
    \rule{0pt}{0.8\normalbaselineskip}
    MINT           & 0.0\% & 0.0\%  & 0.0\% \\ 
    MINT+RFM40     & 0.0\% & 0.0\%  & 0.0\% \\ 
    RrIDE+RFM16     & 1.3\% & 2.1\%  & 1.3\% \\ \hline 
  \end{tabular}
   \vspace{-0.1in}
  \end{footnotesize}
\end{table}
}

\subsection{Comparison with Memory-Controller-Based PARA}

\begin{figure}[!htb]
    \centering
    \vspace{-0.1 in}
\includegraphics[width=3.4in]{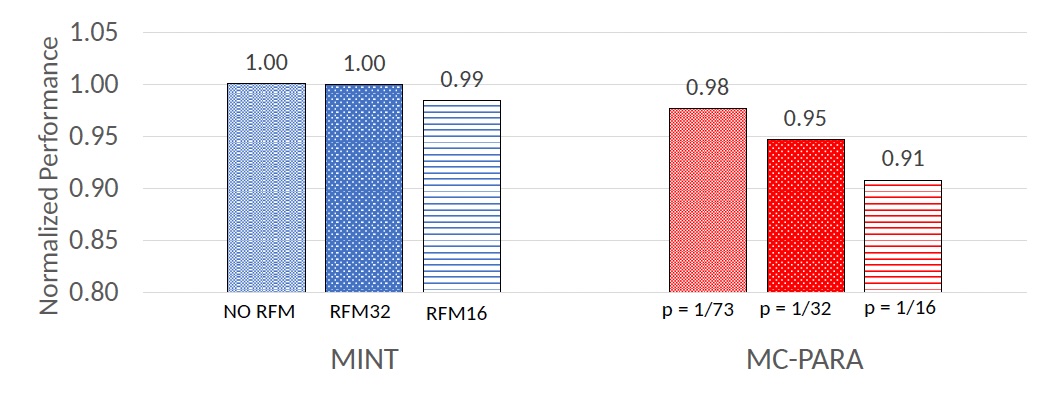}
\vspace{-0.15 in}
    \caption{\textcolor{black}{Performance of MINT and MC-PARA. MC-PARA uses D-RFM which causes all mitigations to be blocking and incurs significant slowdowns.}}
    \vspace{-0.15 in}
    \label{fig:paraperf}
\end{figure}

\textcolor{black}{We compare MINT with PARA implemented on the MC-side (MC-PARA).} Figure~\ref{fig:paraperf} shows the relative performance of MC-PARA and MINT tuned for similar MinTRH.  MINT can do mitigations transparently (within REF) and incur RFM overheads only when ACT count is greater than RFMTH.  MC-PARA relies on DRFM, which means all mitigations block the bank from service.  On average, MC-PARA incurs slowdown from 2-9\%, whereas MINT remains about 1\%. 

\ignore{

DMQ is consulted once per tREFI, and only when refreshes are paused (and all activation slots are used).  The 

MINT incurs energy overheads due to mitigative refreshes, \textcolor{black}{DMQ activity,} and the RNG being consulted on each tREFI. 
Table~\ref{table:energy} shows the average energy overheads for MINT and MINT+RFM using a 7-bit TRNG ~\cite{katz2008robust, yu2019survey} which 
consumes 0.09mW leakage power \textcolor{black}{and a 4-entry DMQ which consumes 0.001mW leakage power} (in 10nm). \textcolor{black}{These structures have negligible impact on overall power/energy since they have five to six orders lower power than the DRAM power}. \textcolor{black}{Accounting for leakage and dynamic energy of the DMQ and PRNG}, MINT incurs only 1\%-3\% extra energy
over an idealized baseline that incurs no energy overheads for mitigation.  MINT increases ACT energy by 6\%--25\%, however, as activation energy is only 13\% of the total energy, its impact on overall energy is small.

}

\newpage
\section{Related Work}

\ignore{
We have previously discussed closely related works including TRR, DSAC, PAT, ProTRR, Mithril, PARFM, and PARA. Now, we describe other related works.
}

\color{black}
\vspace{0.05 in}

\noindent{\bf{Per-Row-Activation-Counting (PRAC):}} Concurrent to our submission, JEDEC announced~\cite{JEDEC-PRAC-News} an update to DDR5 specifications with {\em PRAC}~\cite{JEDEC-PRAC}.  PRAC extends the DRAM array to have a counter and uses that counter to track activations. The timing of memory operations (e.g. tRC) is changed to perform a read-modify-write of the counter for each activation.  While PRAC is a principled defense for Rowhammer, we note that PRAC is an optional feature and DRAM manufacturers are not required to support PRAC. A recent Hynix~\cite{HynixRH} design showed that the area overhead of supporting per-row counters is approximately 9\%. Furthermore, PRAC specifications change the tRC timings from the current 46ns-48ns to 52ns (almost 10\% higher). DRAM industry is extremely cost-sensitive. If there is a secure low-cost method to mitigate Rowhammer, then the memory companies can avoid the significant area, power, and timing overheads of PRAC, and choose the low-cost alternative.  MINT offers such an alternative. Our analysis with feinting-style attacks shows that the threshold of MINT is within 2x of Per-Row-Counter-Table (PRCT).

\ignore{
{Panopticon}~\cite{bennett2021panopticon} stores per-row counters within the DRAM array. These designs change the DRAM array, need to do counter-update after each activation, and incur area overheads. Our low-cost design offers a MinTRH-D within 2x of an idealized design that stores one-counter per row.
}

\color{black}

\vspace{0.05 in}
\noindent{\bf Other One-Counter-Per-Row Designs:} 
 {CRA}~\cite{kim2014architectural} and {Hydra}~\cite{qureshi2022hydra} keep the counters in DRAM and use caches or filters to reduce the DRAM lookups for counters. However, they can incur large slowdowns for the worst-case patterns.

\vspace{0.05 in}
\noindent{\bf Efficient-Counters:} Several proposals reduce the SRAM overhead of aggressor-row tracking. Table~\ref{table:CompareRelatedWork} compares the per-bank SRAM overheads of { Graphene}~\cite{park2020graphene}. MINT has significantly lower SRAM overheads, especially at low TRH.

  \begin{table}[htb]
  \vspace{-0.1 in}
  \centering
  \begin{footnotesize}
 \caption{Per-Bank SRAM Overhead of Trackers (per-rank will be 32x) }
 \vspace{-0.1 in}
  \label{table:CompareRelatedWork}
  \footnotesize{
  \begin{tabular}{ccc}
    \hline
    \textbf{Name} & \textbf{Device TRH-D=3K} & \textbf{Device TRH-D=300}  \\ \hline
    
    Graphene& 56.5 KB & 565 KB \\ 

    \textbf{MINT+DMQ}&  \textbf{15 bytes}  & \textbf{15 bytes} \\ \hline
  \end{tabular}
  }
  \end{footnotesize}
\end{table}

\vspace{0.05 in}
\noindent{\bf Secure Low-Cost Tracker:} Our recent work proposes {\em PrIDE}~\cite{PRIDE}, a secure low-cost in-DRAM tracker.  The InDRAM-PARA design we discuss in this paper is equivalent to single-entry PrIDE. 
PrIDE uses a 4-entry FIFO to reduce the {\em loss probability} (from 63\% to 10\%) but suffers from {\em Tardiness}. In the PrIDE terminology, MINT has zero loss-probability and zero Tardiness (pattern-2).  The MinTRH-D of PrIDE is 1750 (25\% higher than MINT). Refresh postponement causes the MinTRH-D of PrIDE to increase to 6.5K. PrIDE with DMQ has a MinTRH-D of 1900 (28\% higher than MINT+DMQ).

\vspace{0.05 in}
\noindent{\bf Mitigating-Actions:} 
For MINT, we use victim refresh for mitigation.  {RRS}~\cite{saileshwar2022RRS}, {AQUA}~\cite{AQUA},  {SRS}~\cite{SRS}, {SHADOW}~\cite{ShadowHPCA23} perform mitigation with row migration, whereas, {Blockhammer}~\cite{yauglikcci2021blockhammer} uses rate limits. {REGA}~\cite{REGA_SP23} changes the DRAM circuitry to provide mitigating refresh on each demand activation. HiRA~\cite{HIRA} changes the interface to allow multiple activations per bank. MINT avoids changes to DRAM array and interface.  

\vspace{0.05 in}
\noindent{\bf ECC-Codes:} 
{SafeGuard}~\cite{ali2022safeguard}, {CSI-RH}~\cite{csi},  {PT-Guard}~\cite{DSN23_PTGuard}, and Cube~\cite{twobirds}  use ECC codes to tolerate Rowhammer failures,  however, uncorrectable failures can still cause data loss.


\section{Conclusion}

Current in-DRAM trackers for Rowhammer mitigation are either insecure or require prohibitively-high cost.   This paper develops {\em Minimalist In-DRAM Tracker (MINT)} to provide secure Rowhammer mitigation with a single entry.  The key insight that enables MINT is that instead of selecting the aggressor row based on the past or the current, MINT selects one in the future (a random row in the upcoming interval). We also study the compatibility of low-cost trackers with refresh postponement, and propose {\em Delayed Mitigation Queue (DMQ)} as a generalized solution. We show that MINT can securely protect devices with a double-sided TRH of 1482 and as low as 356 when combined with RFM. The storage, performance, and energy overheads of MINT are negligible.


 \color{black}

\section*{Appendix A: Impact of MaxACT}

One of the key parameters that determines the efficacy of in-DRAM trackers is the mitigation rate.  Our default mitigation rate is 1 aggressor row per tREFI. Given our default timing parameters, we can have a maximum activations (MaxACT) of 73 per tREFI. In this section, we consider how the variation in MaxACT across different vendors and specifications can impact the threshold of both MINT and InDRAM-PARA.

JEDEC specifies a range of memory timings and memory companies decide which specifications to support.  For example, for DDR5, JEDEC specifies 11 data transfer rates (DDR5-3200 to DDR5-7200, once every 400), and for each rate, they specify four {\em speed-bins} (A, AN, B, BN).  Across all these 44 specifications, the minimum tRC is 46ns and maximum tRC is 49.5ns (52ns for revised specs with PRAC). For DDR5, tREFW=32ms, and tRFC=350ns or 410ns. Thus, MaxACT can range from 67.2 to 77.2 for the entire DDR5 specs.

\begin{figure}[!htb]
    \centering
\includegraphics[width= 3.4 in]{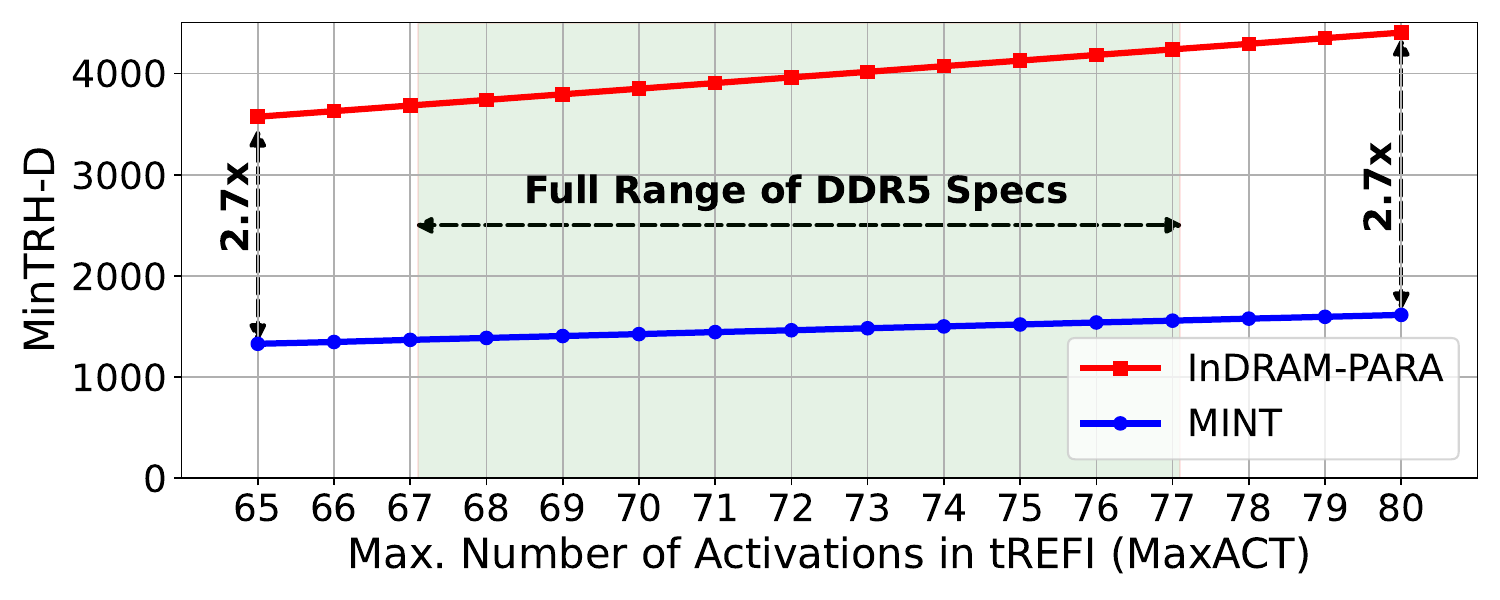}
\vspace{-0.1 in}
    \caption{Impact of varying the maximum number of activations (MaxACT) within tREFI for InDRAM-PARA and MINT. MINT continues to have 2.7x lower MinTRH-D than InDRAM-PARA across the entire range of MaxACT.}
    \vspace{-0.15 in}
    \label{fig:trefi}
\end{figure}

Figure~\ref{fig:trefi} shows the MinTRH-D supported by MINT and InDRAM-PARA as the MaxACT is varied from 65 to 80.  The viable range of MaxACT for the DDR5 specifications is highlighted in green. For our evaluations, we use a MaxACT value of 73.  The MinTRH-D of both MINT and In-DRAM paper increases almost in direct proportion to MaxACT. This is expected as the probability of mitigation reduces as the number of activation slots within the tREFI window increases.  However, it can be noted that the relative difference between MINT and InDRAM-PARA remains at 2.7x throughout the DDR5 range (and even outside). Thus, the advantage provided by MINT is not limited to a specific choice of MaxACT. 

 \color{black}

\section*{Appendix B: Adaptive Attacks on MINT+DMQ}

\begin{figure}[!htb]
    \centering
       \vspace{-0.05 in}
\includegraphics[width= 3.25 in]{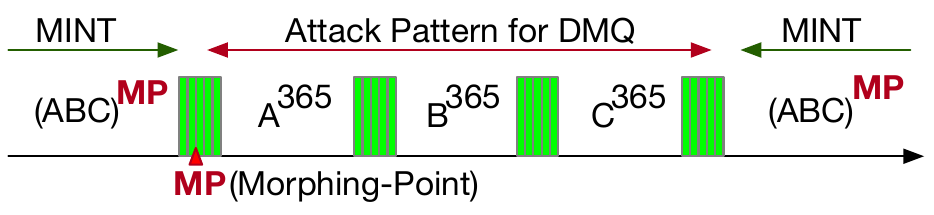}
\vspace{-0.1 in}
    \caption{ADA morphs the pattern from MINT to DMQ at morphing-point. We show 3 rows for simplicity, MINT pattern has 73 lines.  }
   \vspace{-0.1 in}
    \label{fig:ada}
\end{figure}

\noindent{\bf{The Pattern:}} The best attack for MINT is to activate a given aggressor row within tREFI only once (to evade selection).  The best attack for DMQ is to activate the same row repeatedly, as it allows more activations on the selected row while waiting in the DMQ for mitigation.  Therefore, using a single pattern is not ideal for the MINT+DMQ. We develop an {Adaptive Attack (ADA) on DMQ} that changes the pattern from what is optimal for MINT (pattern-2) to what is optimal for DMQ (repeated activations) at a predefined {\em morphing-point (MP)}. Figure~\ref{fig:ada} shows an overview of  ADA.

As refresh postponement can allow 365 activations on a given row, this switching can allow ADA to cause 365 activations on an aggressor row (without any intervening mitigation), however, after ADA, this row is guaranteed to get mitigated.  Thus, if the row had $A$ activations at the morphing-point, then ADA can cause the row to have $A+365$ activations.

\begin{figure}[!htb]
    \centering
       \vspace{-0.05 in}
\includegraphics[width= 3in]{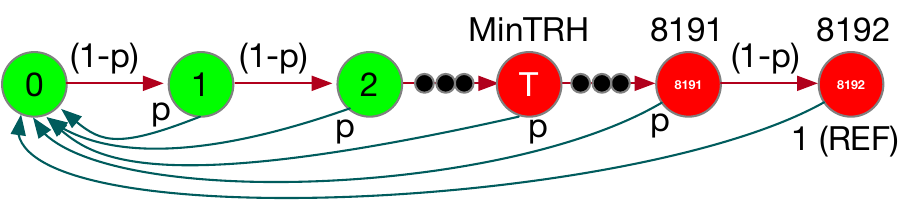}
\vspace{-0.1 in}
    \caption{Markov-Chain model for activation count of a row.  Each state denote the activation counts since last mitigation or refresh. The parameter {\em p} denotes the probability of mitigation. Red/Green denotes states with/without an error.}
  \vspace{-0.05 in}
    \label{fig:markov}
\end{figure}

\vspace{0.05 in}
\noindent{\bf The Model for Threshold:} For ADA to be effective it must find a row with at-least (MinTRH - 365) activations, otherwise the row is guaranteed to get mitigated before reaching MinTRH. Similarly, if the row already has MinTRH, it fails without requiring ADA.  As the attacker does not know the activation-counts of a given row within the given tREFW window, the key decision for ADA is to set the morphing-point (in terms of tREFI), with the useful range from (MinTRH-365) to (8192 - tREFI).  To compute the probability of finding a given row with a given activation count (A) at a given tREFI interval, we use a Markov-Chain, as shown in Figure~\ref{fig:markov}. For pattern-2, the row can have an activation-count from 1 to 8192 (states above MinTRH indicate failure with MINT). We determine the probability of finding a row with A activations and then increase it by 365 due to ADA. We use these probabilities to compute the $P_{REFW}$, MTTF, and MinTRH.

\begin{figure}[!htb]
    \centering
\includegraphics[width= 3.4in]{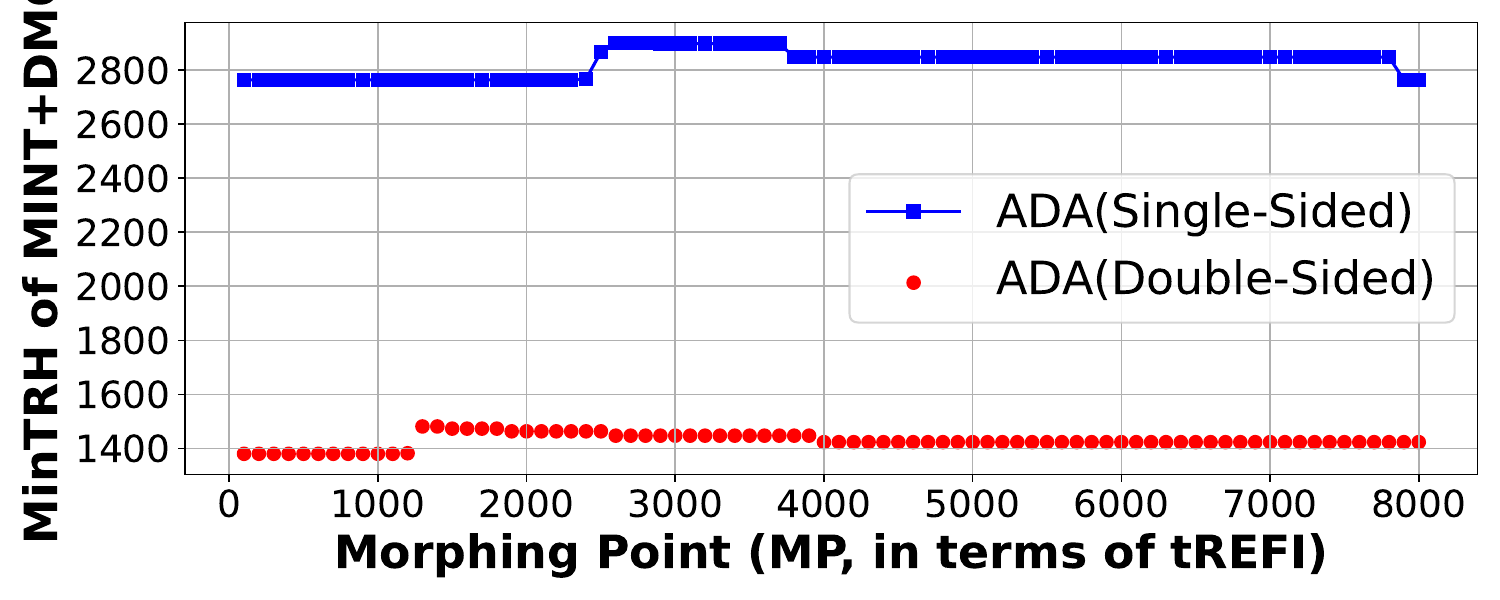}
    \caption{Threshold of MINT+DMQ with ADA, as the Morphing-Point (MP) is varied. ADA has MinTRH of 2899 and MinTRH-D of 1482. The double-sided attack has shorter period for DMQ attack and an earlier MP.  }
    \label{fig:trhdmq}
\end{figure}

\vspace{0.05 in}
\noindent{\bf The Impact:} Figure~\ref{fig:trhdmq} shows the MinTRH of ADA as the morphing-point is varied from 500 to 8000. We evaluate both single-sided and double-sided version. For single-sided version, ADA starts to become effective only after MP of 2400 (before this the MinTRH remains 2763, same as without ADA), and provides the highest MinTRH of 2899 at MP between 2533-3730, beyond this the MinTRH drops to 2847. The reason for this behavior is a smaller MP allows for the attack to be repeated multiple times within the same tREFW. For the double-sided pattern, the attack starts to become effective after MP of 1200, and provides the highest MinTRH-D of 1282 between MP of 1299 and 1456, after which MinTRH-D reduces to 1474. The earlier success point for double-sided attack is because MinTRH-D (without ADA) is much lower than MinTRH (without ADA). Overall, this analysis shows  that under adaptive attacks, {\bf MinTRH-D of MINT+DMQ is 1482}.

\section*{Appendix C: Tolerating Row-Press with MINT}

Row-Press~\cite{rowpress} is a new vulnerability that arises when a row is kept open for a long time.  The charge in the nearby rows continue to slowly leak through the bit lines.  Each round of Row-Press incurs one activation and keeps the row open for a period of tON (tON can be between tRAS and 5*tREFI).  Due to the extra charge leaked during tON, Row-Press can cause bit flips in much fewer activations than TRH. 

Our concurrent work, ImPress~\cite{IMPRESS}, enables in-DRAM trackers to mitigate Row-Press without affecting the tolerated TRH. The key idea is to convert the row open time into an equivalent number of activations (EACT) for Rowhammer mitigation.   Thus, rows that are kept open for a longer time have higher EACT and therefore a higher rate of mitigation.  EACT is given by Equations~\ref{eq:eact}.
\begin{equation}
\label{eq:eact}
    EACT=(tON+tPRE)/tRC
\end{equation}
ImPress requires a timer to track tON.  The division (with tRC) is implemented with a shift operation. EACT can have up to 7-bits of fractional part. To tolerate Row-Press with MINT, we must change the 7-bit CAN register to a fixed-point register (7+7=14 bits). For each activation, CAN is incremented by an amount equal to EACT.  When the value of CAN crosses SAN, the row causing the activation is stored in SAR.  MINT combined with ImPress can tolerate Row-Press without affecting the MinTRH.  With ImPress, the total storage overhead increases from 15 to 17 bytes per bank. 

\section*{Acknowledgements}

We thank the reviewers of MICRO-2024 for their feedback. This research was supported, in part,  by a gift from Intel.
\bibliographystyle{IEEEtranS}
\bibliography{refs}


\end{document}